\documentclass[reprint, aps, pra, footinbib, letterpaper, superscriptaddress]{revtex4-1}

\usepackage[T1]{fontenc} 

\usepackage{amsmath,color}
\usepackage{amssymb}
\usepackage{amsfonts}
\usepackage{amsthm}
\usepackage{mathtools}

\usepackage[title]{appendix}

\usepackage{algorithm}
\usepackage{algpseudocode}
\usepackage{dsfont}

\DeclarePairedDelimiter\bra{\langle}{\rvert}
\DeclarePairedDelimiter\ket{\lvert}{\rangle}
\DeclarePairedDelimiterX\braket[2]{\langle}{\rangle}{#1 \delimsize\vert #2}

\newcommand{\vJ}{\mathbf{J}}

\newcommand{\vE}{\mathbf{E}}

\newcommand{\vB}{\mathbf{B}}

\newcommand{\ve}{\mathbf{e}}
\newcommand{\vnabla}{\boldsymbol{\nabla}}

\newcommand{\kFGR}{k_{\text{FGR}}}

\newcommand{\he}{\mathbf{e}}
\newcommand{\hH}{\hat{H}}

\newcommand{\hV}{\hat{V}}
\newcommand{\hsigma}{\hat{\sigma}}

\newcommand{\hP}{\hat{\boldsymbol{\mathcal{P}}}}

\newcommand{\hrho}{\hat{\rho}}

\newcommand{\tr}[1]{\text{Tr}\left(#1\right)}

\newcommand{\avg}[1]{\left\langle #1\right\rangle}

\newcommand{\red}[1]{{\color{black} #1}}
\newcommand{\blue}[1]{{\color{black} #1}}

\begin{document}

	\title{Quasiclassical Modeling of Cavity Quantum Electrodynamics}%
	
	\author{Tao E. Li}%
	\email{taoli@sas.upenn.edu}
	\affiliation{Department of Chemistry, University of Pennsylvania, Philadelphia, Pennsylvania 19104, USA}

	\author{Hsing-Ta Chen}
	\affiliation{Department of Chemistry, University of Pennsylvania, Philadelphia, Pennsylvania 19104, USA}
	
	\author{Abraham Nitzan} 
	\affiliation{Department of Chemistry, University of Pennsylvania, Philadelphia, Pennsylvania 19104, USA}

	\author{Joseph E. Subotnik}
	\email{subotnik@sas.upenn.edu}
	\affiliation{Department of Chemistry, University of Pennsylvania, Philadelphia, Pennsylvania 19104, USA}

	\begin{abstract}
		 We model  a collection of $N$ two-level systems (TLSs) coupled to a multimode cavity via  Meyer-Miller-Stock-Thoss (MMST) dynamics, sampling both  electronic and photonic zero-point energies  (ZPEs) and propagating independent trajectories in Wigner phase space. By investigating the ground state stability of a single TLS, we use MMST dynamics to separately study both electronic ZPE effects (which  would naively lead to the breakdown of the electronic ground state) as well as photonic ZPE effects (which would naively lead to spontaneous absorption). By contrast, including both effects (i.e., sampling both electronic and photonic ZPEs) leads to the dynamical stability of the electronic ground state.  Therefore, MMST dynamics provide a practical way to identify the contributions of self-interaction and vacuum fluctuations. More importantly, we find that MMST dynamics can predict accurate quantum dynamics for both electronic populations and \red{electromagnetic} field intensity \red{in the high saturation limit}.  For a single TLS in a cavity, MMST dynamics correctly predict the initial exponential decay of spontaneous emission, Poincar\'e recurrences, and the positional dependence of a spontaneous emission rate. For an array of $N$ equally spaced TLSs with only one TLS excited initially, MMST dynamics  correctly predict the modification of spontaneous emission rate as a function of the spacing between TLSs. Finally, MMST dynamics also correctly model Dicke's superradiance and subradiance (i.e., the dynamics when all TLSs are excited initially) including the correct quantum statistics for the delay time (as found by counting trajectories, for which a full quantum simulation is hard to achieve). Therefore, this work raises the possibility of simulating large-scale collective light-matter interactions with methods beyond mean-field theory.
	\end{abstract}

	\maketitle
	
	
	\section{Introduction}\label{sec:intro}
	In confined geometries such as nanoscale cavities,  the quantum nature of photons can strongly modify the properties of atoms and molecules, including the control of spontaneous emission rates\cite{Purcell1946,Jacob2012,Buzek1999,Albrecht2019}, frequency splitting of the absorption spectrum due to strong light-matter coupling\cite{Yoshie2004,Zengin2015}, and changes in chemical reaction landscapes by forming hybrid light-matter states (molecular polaritons)\cite{Hutchison2012,Herrera2016,Munkhbat2018,Galego2019,Campos-Gonzalez-Angulo2019,Mandal2019}. In the field of cavity quantum electrodynamics (cQED), theorists traditionally describe these phenomena by adapting simplified quantum models, such as the Jaynes--Cummings (JC) model\cite{Jaynes1963} [i.e., a two-level system (TLS) coupled to a single cavity photon mode],  the Tavis--Cummings (TC) model\cite{Tavis1968,Tavis1969} (i.e., $N$ TLSs coupled to a single photon mode),  or the Weisskopf-Wigner model\cite{weisskopf1930z} (i.e., a TLS coupled to $M$ photon modes within the context of a  single excitation manifold). With these simplified quantum models, many exciting quantum phenomena can be studied analytically\cite{Meystre2007}. 
	
	Going beyond simplified models,  due to the increasing complexities of the full quantum light-matter Hamiltonian when realistic atoms or molecules are considered, finding analytical solutions becomes increasingly difficult\cite{Braak2011}. By contrast, numerically propagating both electronic and photonic degrees of freedom  (DoFs) together becomes a good choice. For such a computational problem in cQED, one approach is to keep all variables quantum-mechanically and seek for approximate quantum solutions (often with the spirit of mean field theory); in principle, quantum-electrodynamical density functional theory (QEDFT)\cite{Pellegrini2015,Flick2017} should be exact if one knew the correct exchange-correlation functional. An alternative is to seek a semiclassical approximation whereby some DoFs (e.g., the electrons) are kept quantum-mechanically (and propagated exactly) but other DoFs (e.g., the photons) are propagated classically (again exactly).
	The most popular such approach today is the  coupled Maxwell-Schr\"odinger equations\cite{Crisp1969,Lorin2007,Sukharev2011,Gao2012,Li2016}.

	Now, the major problem underlying the coupled Maxwell-Schr\"odinger equations is that all quantum effects of the \red{electromagnetic (EM)} fields are completely ignored, and thus, this method cannot be used when the quantum dynamics of the radiation field are important, e.g., \red{in the high saturation limit\cite{Crisp1969,Li2018Tradeoff,Li2019Hamiltonians,Chen2019Mollow} or} in cavities. Recently, attempts have been made to include quantum EM-field effects even when evolving classical EM fields\cite{Li2018Spontaneous,Hoffmann2018}. For example, in the recently proposed Ehrenfest+R approach\cite{Chen2018Spontaneous,Chen2019Mollow,Li2019Stimulated,Chen2018Raman}, our research group included vacuum fluctuations by propagating a swarm of augmented Maxwell-Schr\"odinger equations. Nevertheless,  because Ehrenfest+R was developed in  free space,  the performance of the method in cavities is unknown. As another attempt, Hoffmann \textit{et al} have proposed a multi-trajectory Ehrenfest approach in cavities\cite{Hoffmann2019}. In this approach, the electronic DoFs are evolved quantum-mechanically and the EM fields are propagated classically with initial conditions that include sampling the photonic zero-point energy (ZPE) in Wigner phase space. While this approach can predict some quantum effects of spontaneous emission, the agreement with quantum solutions is still far from quantitatively accurate. For a benchmark study of semiclassical approaches applied to a single two- or three-level system in a cavity, see the recent work of Hoffmann \textit{et al}\cite{Hoffmann2019Benchmark}.

	In this paper, we will analyze yet another, slightly more rigorous semiclassical approach for cQED dynamics --- the Meyer-Miller-Stock-Thoss (MMST) approach\cite{Meyera1979,Stock1997}, a method which was originally developed for  coupled electron-nuclear dynamics [and is also known as the Poisson bracket mapping equation (PBME)\cite{Kapral1999,kim2008quantum,nassimi2010analysis,kim2014improving,kim2014improving2}]. The basic philosophy underlying this method is to  map all \red{electronic} DoFs to Wigner phase space \red{as harmonic oscillators}.
	\red{Note that this mapping is exact, as shown by Stock and Thoss\cite{Stock1997}.  For instance, if there is an electronic state $\ket{k}$, one simply replaces $\ket{k}$ by a creation operator $\hat{a}_k^{\dagger}$, and similarly one replaces $\bra{k}$  by an annihilation operator $\hat{a}_k$.  Then, one uses the canonical relationship $\hat{a}_k = (\hat{x}_k+ i \hat{p}_k)/\sqrt{2}$ to rewrite the electronic Hamiltonian as a function of $\hat{x}_k$ and $\hat{p}_k$.  At this point, one sacrifices exactness and invokes the classical ansatz of treating $\hat{x}_k$ and $\hat{p}_k$ classically (with initial values sampled from a distribution, known as the initial value representation).  The resulting, independent quasiclassical trajectories can be used to recover the total wave packet evolution in phase space, and effectively reduce to mean-field Ehrenfest dynamics with only one caveat: the electronic (and nuclear) degrees of freedom are given ZPEs.  Thus, for example, the electronic wave function is not formally normalized to one.  For a  more detailed explanation of this method, see Sec. \ref{sec:method}. This flavor of initial value representations has been used for many years in theoretical chemistry, largely going back to the work of Miller and co-workers\cite{Meyera1979,Tao2010}.  
	Recent work by Cotton and Miller has shown that this flavor of dynamics can perform quite well for the spin-boson model\cite{Cotton2013,Cotton2016,Cotton2019}, which is effectively the exact same Hamiltonian as Eq. \eqref{eq:H_tot_QED} when only one TLS is considered.
	For coupled electron-photonic systems, because the photonic DoFs are exactly harmonic which is optimal for mean-field dynamics, quasiclassical MMST dynamics should behave quite well.    Thus, altogether, quasiclassical MMST dynamics should be advantageous relative to the traditional  coupled  Maxwell-Schr\"odinger equations for solving coupled electronic-photonic dynamics.}

	Indeed, by studying $N$ TLSs coupled to a multimode cavity, we will show that MMST dynamics provide an intuitive physical interpretation for a cQED problem. Most importantly, we will show how MMST dynamics can sometimes predict accurate collective quantum dynamics in cavities including superradiance and subradiance phenomena.
	
	This paper is organized as follows. In Sec. \ref{sec:method} we introduce the quantum model and MMST dynamics. In Sec. \ref{sec:simulation_details} we list the simulation details. In Sec. \ref{sec:results} we present the results for $N$ TLSs coupled to a multimode cavity. We conclude in Sec. \ref{sec:disscussion_conclusion}. As far as notation is considered, we let Roman character $j$ denotes cavity modes and Greek character $\alpha$ denotes TLSs.

	\section{Method}\label{sec:method}
	\subsection{Quantum Dynamics}
	For a collection of $N$ TLSs interacting with a multimode cavity,  the full quantum Hamiltonian reads:
		\begin{subequations}
		\label{eq:H_tot_QED}
		\begin{equation}
		\label{eq:H_tot_QED_1}
				\begin{aligned}
		\hH &=  \sum_{\alpha= 1}^{N} \frac{1}{2}\hbar\omega_0\hsigma_z^{(\alpha)} + \sum_{j=1}^{M} \hbar\omega_j \left(\hat{a}^{\dagger}_{j}\hat{a}_{j} + \frac{1}{2}\right) \\
		&- \sum_{\alpha =1}^{N}\sum_{j=1}^{M}\red{\hbar} g_j^{(\alpha)}\left(\hat{a}^{\dagger}_j + \hat{a}_j\right)\left(\hsigma_{+}^{(\alpha)} + \hsigma_{-}^{(\alpha)}\right)
		\end{aligned}
		\end{equation}
		Here,  $\hbar\omega_0$ denotes the energy gap between the ground state ($\ket{\alpha g}$) and excited state ($\ket{\alpha e}$) for each TLS, 
		$\hsigma_z^{(\alpha)} \equiv \ket{\alpha e}\bra{\alpha e} - \ket{\alpha g}\bra{\alpha g}$, $\hsigma_{+}^{(\alpha)} \equiv \ket{\alpha e}\bra{\alpha g}$, and $\hsigma_{-}^{(\alpha)} \equiv \ket{\alpha g}\bra{\alpha e}$.
		$\hat{a}^{\dagger}_{j}$ and $\hat{a}_{j}$ denote the creation and annihilation operators for the $j$-th photon mode with energy $\hbar\omega_j$, where $\omega_{j} = \frac{j\pi c}{L}$ and $L$ denotes the length of cavity. For simplicity, we assume that each photon couples to TLSs via only a single polarization direction, so we do not sum over two polarization vectors here. We also truncate the number of photon modes to a finite number ($M$ modes) to facilitate numerical calculations.
		Finally, the position-dependent coupling constant $g_j^{(\alpha)}$ is defined as
		\begin{align}\label{eq:gj}
		g_j^{(\alpha)} = \sqrt{\frac{\omega_j}{\hbar\epsilon_0 L}}\mu_{ge}^{(\alpha)}\sin(k_j r_\alpha)
		\end{align}
		where $\mu_{ge}^{(\alpha)}$ denotes the magnitude of transition dipole moment for the TLS located at $r_\alpha$, and $k_j = \frac{\omega_j}{c} =  \frac{j\pi}{L}$. 
		\end{subequations}

		 In order to capture the real-time dynamics for the coupled electron-photonic system, one can evolve the time-dependent Schr\"odinger equation:
		\begin{equation}
		i\hbar \frac{\partial}{\partial t} \ket{\Psi(t)}  = \hH \ket{\Psi(t)} 
		\end{equation}
		where $\ket{\Psi(t)}$ denotes the wave function for the coupled electron-photonic system. 
		Practically, in order to numerically propagate the Schr\"odinger equation, one needs to choose a truncated  basis which includes up to $D$ excitation(s). If $D = 1$, the truncated basis is the configuration interaction singles (CIS) basis:
		\begin{equation}\label{eq:phi_CIS}
		\begin{aligned}
		\ket{\Psi_{\text{CIS}}(t)} &= c_0(t)\ket{\vec{g}}\ket{\vec{0}}
		+ \sum_{\alpha= 1}^{N}c_\alpha(t)\ket{e_\alpha}\ket{\vec{0}} \\
		&+ \sum_{j = 1}^{M}d_j(t)\ket{\vec{g}}\ket{1_j}
		\end{aligned}
		\end{equation}
		Here, $\ket{\vec{g}}$ denotes a wave function for which all of the TLSs are  in ground state, and $\ket{e_\alpha}$ denotes a wave function for which only the $\alpha$-th TLS is excited. Similarly, $\ket{\vec{0}}$ denotes a wave function for which all photon modes are in the ground state, and $\ket{1_j}$ denotes a wave function for which the $j$-th photon mode has one photon but all other modes are in their respective ground states (with zero photons). 
		\red{Note that the CIS approximation implies that a rotating-wave approximation is taken in Eq. \eqref{eq:H_tot_QED_1}, i.e., $\left(\hat{a}^{\dagger}_j + \hat{a}_j\right)\left(\hsigma_{+}^{(\alpha)} + \hsigma_{-}^{(\alpha)}\right) \approx \hat{a}^{\dagger}_j\hsigma_{-}^{(\alpha)} + \hat{a}_j\hsigma_{+}^{(\alpha)}$, and we ignore all effects due to the counter rotating-wave terms.}
		With this  CIS approximation, the truncated wave function has dimension $1 + N + M$. Generally, for $D \geq 2$, the dimensionality of the truncated wave function grows uncontrollably with increasing $D$, which prohibits real-time simulations for highly excited systems (e.g., Dicke's superradiance). 
		
		Now, during a simulation, we will be interested in the expectation values of various key operators.  To that end, the excited state population for each TLS can be calculated by evaluating $\rho_{ee}^{(\alpha)}(t) = \left \langle\Psi(t)\bigg|\hrho_{ee}^{(\alpha)} \bigg|\Psi(t)\right\rangle$, where
		\begin{equation}\label{eq:population_operator}
		\hrho_{ee}^{(\alpha)} = \ket{\alpha e}\bra{\alpha e}
		\end{equation}
		For the photonic part, the  E-field and B-field operators read:
		\begin{subequations}
		\begin{align}
		\label{eq:observation_operator_QED_hEperp}
		\hat{E}(r) &= \sum_{j} \varepsilon_j \left(\hat{a}^{\dagger}_j + \hat{a}_j\right)\sin(k_j r) \\
		\hat{B}(r) &= \sum_{j} \frac{i}{c}\varepsilon_j \left(\hat{a}^{\dagger}_j - \hat{a}_j\right)\cos(k_j r)
		\end{align}
		\end{subequations}
		where $\varepsilon_j = \sqrt{\hbar\omega_j / \epsilon_0 L}$. We will also be interested in the normal-ordered field intensity operator:
		\begin{equation}\label{eq:intensity_operator}
		\begin{aligned}
		\hat{I}(r) &= :\epsilon_0\hat{E}^2(r) : \\
		&=\epsilon_0\hat{E}^2(r) - \epsilon_0\sum_{j}\varepsilon_j^2 \sin^2(k_j r) 
		\end{aligned}
		\end{equation}
		Here, the colons ($::$) indicate the normal ordering, and $:\hat{a}^{\dagger}_j\hat{a}_{j}: = :\hat{a}_j\hat{a}^{\dagger}_{j}: = \hat{a}^{\dagger}_j\hat{a}_{j}$. The normal-ordered intensity excludes the effect of photonic zero point energies (ZPE), i.e., if the photonic field is the vacuum field ($\Psi(t) = \ket{\vec{g}}\ket{\vec{0}}$), then $ \avg{\Psi(t)\bigg|\hat{I}(r)\bigg | \Psi(t)} = 0$ everywhere.
		
	\subsection{Quasiclassical Meyer-Miller-Stock-Thoss (MMST) Dynamics}
	
		For solving a cQED problem, much like any semiclassical problem, it is standard to directly evolve the Schr\"odinger equation  for the quantum subsystem.
		Alternatively, we can also take another strategy: mapping the full quantum Hamiltonian into phase space, and then recovering quantum dynamics by sampling quasiclassical trajectories in  phase space. One approach of this kind is the Meyer-Miller-Stock-Thoss (MMST) approach\cite{Meyera1979,Stock1997}. While MMST dynamics were developed to solve  coupled electron-nuclear dynamics,  this approach can also describe the coupled electron-photonic dynamics very well; e.g., see  the recent work by Hoffmann \textit{et al}\cite{Hoffmann2019Benchmark} \red{(in which they refer to this approach as the linearized semiclassical dynamics)}. For the sake of clarity, we will now provide a brief review.
	
	\subsubsection{MMST Mapping}
	
		The first step of MMST dynamics is the MMST mapping\cite{Meyera1979,Stock1997}, which provides a systematic and exact way to map a coupled electron-photonic  Hamiltonian onto a set of quantum harmonic oscillators. Now, mapping photonic DoFs to harmonic oscillators is trivial: one just needs to replace $\hat{a}^{\dagger}_j$ and $\hat{a}_j$ with Cartesian operators $\hat{X}_j$ and $\hat{P}_j$ using the following identities:
		\begin{subequations}\label{eq:XP_photon}
			\begin{align}
			\label{eq:XP_photon_X}
			\hat{X}_j &= \sqrt{\frac{\hbar}{2\omega_j}}\left(\hat{a}^{\dagger}_j + \hat{a}_j\right) \\
			\hat{P}_j &= i\sqrt{\frac{\hbar\omega_j}{2}}\left(\hat{a}^{\dagger}_j - \hat{a}_j\right)
			\end{align}
		\end{subequations}
		For the electronic DoFs, the MMST mapping states that the electronic operators can be written as a string of annihilation and creation operators:
		\begin{equation}
		\ket{\alpha k}\bra{\alpha l} \rightarrow \hat{a}_{\alpha k}^{\dagger}\hat{a}_{\alpha l}
		\end{equation}
		where $k, l = e, g$. By further writing $\hat{a}_{\alpha k} = \frac{1}{\sqrt{2}}\left(\hat{x}_{\alpha k} + i\hat{p}_{\alpha k}\right)$, electronic operators can also be mapped to Cartesian coordinate operators, such that
		\begin{subequations}\label{eq:XP_electron}
			\begin{align}
			\ket{\alpha k}\bra{\alpha k} &\rightarrow \frac{1}{2}\left(\hat{x}_{\alpha k}^2 + \hat{p}_{\alpha k}^2 - 1\right) \\
			\ket{\alpha k}\bra{\alpha l} + \ket{\alpha l}\bra{\alpha k}  &\rightarrow \hat{x}_{\alpha k}\hat{x}_{\alpha l}+ \hat{p}_{\alpha k}\hat{p}_{\alpha l} \text{\ \ \ } (k \neq l) \\
			i\left(\ket{\alpha k}\bra{\alpha l} - \ket{\alpha l}\bra{\alpha k}\right)  &\rightarrow \hat{p}_{\alpha k}\hat{x}_{\alpha l}-  \hat{x}_{\alpha k}\hat{p}_{\alpha l} \text{\ \ \ } (k \neq l)
			\end{align}
		\end{subequations}
		By substituting the Cartesian operators  (Eqs. \eqref{eq:XP_photon} and \eqref{eq:XP_electron}) into  the full quantum Hamiltonian (Eq. \eqref{eq:H_tot_QED}), we arrive at the MMST mapping Hamiltonian for the coupled electron-photonic system: 
		\begin{equation}
		\label{eq:H_QED_MMST}
		\begin{aligned}
		\hH &=  \sum_{\alpha = 1}^{N}\sum_{k = g,e} h_{kk}\left(\frac{1}{2}\hat{x}_{\alpha k}^2 + \frac{1}{2}\hat{p}_{\alpha k}^2 - \gamma \right)\\  
		&+ \sum_{j=1}^{M}\frac{1}{2}\left(\hat{P}_j^2 + \omega_j^2 \hat{X}_j^2 \right)\\
		&-  \sum_{\alpha =1}^{N}\sum_{k\neq l = e, g} 
		\hat{h}_{kl}
		 \left(\hat{x}_{\alpha k}\hat{x}_{\alpha l}+ \hat{p}_{\alpha k}\hat{p}_{\alpha l}\right)
		\end{aligned}
		\end{equation}
		 Here,
		 the relevant coefficients are $h_{ee} = \frac{1}{2}\hbar\omega_{0}$, $h_{gg} = - \frac{1}{2}\hbar\omega_{0}$, and $\hat{h}_{eg} = \hat{h}_{ge} = \sum\limits_{j=1}^{M}
		g_j^{(\alpha)}\sqrt{\frac{\omega_{j}}{2\hbar}} \hat{X}_j$, and $\gamma = \frac{1}{2}$ denotes the electronic zero-point energy. Note that the  MMST mapping Hamiltonian (Eq. \eqref{eq:H_QED_MMST}) is equivalent to the original full quantum Hamiltonian  (Eq. \eqref{eq:H_tot_QED}). However, the advantage of the  MMST mapping Hamiltonian is that the Cartesian operators can be easily connected to Wigner phase space, which will facilitate further approximations.
		Henceforward, we will use the notation $\hat{\mathbf{X}} = \{\hat{x}_{\alpha k}, \hat{X}_j\}$ and $\hat{\mathbf{P}} = \{\hat{p}_{\alpha k}, \hat{P}_j\}$ to connote \red{the set of \textit{both} electronic and photonic variables and we will}  refer to  Eq. \eqref{eq:H_QED_MMST} as $\hH\left(\hat{\mathbf{X}}, \hat{\mathbf{P}}\right)$. 
	
	\subsubsection{Wigner Phase Space Dynamics}
		Quantum-mechanically, a quantum density operator $\hat{\rho}(\hat{\mathbf{X}}, \hat{\mathbf{P}}, t)$ exactly describes the state of a quantum system and  obeys the quantum Liouville equation:
		\begin{equation}\label{eq:drhodt_QED_MMST}
		\frac{\partial}{\partial t}\hat{\rho}(\hat{\mathbf{X}}, \hat{\mathbf{P}}, t) = -\frac{i}{\hbar}\left[\hH\left(\hat{\mathbf{X}}, \hat{\mathbf{P}} \right), \hat{\rho}(\hat{\mathbf{X}}, \hat{\mathbf{P}}, t)\right]
		\end{equation}
	where $\hH\left(\hat{\mathbf{X}}, \hat{\mathbf{P}} \right)$ is given in Eq. \eqref{eq:H_QED_MMST}. Let us perform a Wigner transform\cite{Hillery1984,OzorioDeAlmeida1998} of the quantum density operator $\hat{\rho}(\hat{\mathbf{X}}, \hat{\mathbf{P}}, t)$:
	\begin{widetext}
	\begin{equation}
	\rho_{W}(\mathbf{X}, \mathbf{P}, t) = \left(\frac{1}{\pi\hbar}\right)^{F} \int_{-\infty}^{\infty} e^{-2i\mathbf{P}\cdot \mathbf{Y}/\hbar} \avg{\mathbf{X} + \mathbf{Y}\ \bigg|\ \hat{\rho}(\hat{\mathbf{X}}, \hat{\mathbf{P}}, t)\ \bigg|\ \mathbf{X} - \mathbf{Y}}d\mathbf{Y}
	\end{equation}
	\end{widetext}
	where $F$ denotes the total number of DoFs. In this way, the quantum density operator $\hat{\rho}(\hat{\mathbf{X}}, \hat{\mathbf{P}}, t)$ is transformed to a quasiclassical phase space density $\rho_{W}(\mathbf{X}, \mathbf{P}, t)$, \red{where 
	$\mathbf{X} = \{x_{\alpha k}, X_j\}$ and $\mathbf{P} = \{p_{\alpha k}, P_j\}$.}
	 If the equation of motion for $\rho_{W}$ is cut off as first order in $\hbar$\cite{Hillery1984,OzorioDeAlmeida1998}, one recovers the classical Liouville equation:
	\begin{equation}\label{eq:drhodt_Wigner_MMST}
	\frac{\partial}{\partial t}\rho_{W}(\mathbf{X}, \mathbf{P}, t) = - \left\{ \rho_{W}(\mathbf{X}, \mathbf{P}, t), H(\mathbf{X}, \mathbf{P}) \right\}  \ +\ \red{O(\hbar^2)}
	\end{equation}
	Here, $\{\cdots\}$ denotes the Poisson bracket, and the full classical Hamiltonian $H(\mathbf{X}, \mathbf{P})$ reads
	\begin{equation}
	\label{eq:H_Wigner_MMST}
	\begin{aligned}
	H\left(\mathbf{X}, \mathbf{P} \right) = & \sum_{\alpha = 1}^{N}\sum_{k = g,e} h_{kk} \left(\frac{1}{2}x_{\alpha k}^2 + \frac{1}{2}p_{\alpha k}^2 - \gamma \right)  \\
	+& \sum_{j=1}^{M} \frac{1}{2}\left(P_j^2 + \omega_j^2 X_j^2\right) \\  - & \sum_{\alpha =1}^{N}\sum_{k\neq l = e, g} h_{kl} \left(x_{\alpha k}x_{\alpha l}+ p_{\alpha k}p_{\alpha l}\right)
	\end{aligned}
	\end{equation}
	and $h_{kl} = \sum\limits_{j=1}^{M} \red{\hbar}
	g_j^{(\alpha)}\sqrt{\frac{\omega_{j}}{2\hbar}} X_j$.  	
	Within Wigner phase space, the expectation value of a given operator $\hat{A}$ can be calculated by
	\begin{equation}
	\avg{\hat{A}} = \iint d\mathbf{X} d\mathbf{P} \ A_{W}(\mathbf{X}, \mathbf{P})\rho_{W}(\mathbf{X}, \mathbf{P}, t)
	\end{equation}
	where $A_{W}(\mathbf{X}, \mathbf{P})$ denotes the Wigner-Weyl transform of $\hat{A}$:
	\begin{equation}\label{eq:A_W}
	A_W(\mathbf{X}, \mathbf{P}) = 2\int_{-\infty}^{\infty}  e^{2i\mathbf{P}\cdot\mathbf{Y}/\hbar}\avg{\mathbf{X} - \mathbf{Y}\ \bigg | \ \hat{A} \  \bigg|\ \mathbf{X} + \mathbf{Y}}d\mathbf{Y}
	\end{equation}
	\red{In Eq. \eqref{eq:A_W},  $\hat{A}$ can be either an electronic, photonic, or joint electron-photon operator.}

	\subsubsection{Sampling Independent  Trajectories}
	 While directly evolving the classical phase space density $\rho_{W}(\mathbf{X}, \mathbf{P}, t)$ according to Eq. \eqref{eq:drhodt_Wigner_MMST} is computationally expensive when the dimensions of  phase space is large, an efficient way to propagate Eq. \eqref{eq:drhodt_Wigner_MMST} is to propagate \textit{independent} trajectories in phase space, i.e., assuming
	 \begin{equation}\label{eq:rhoW_sample_traj}
	 \rho_{W}(\mathbf{X}, \mathbf{P}, t) \approx \frac{1}{N_{\text{traj}}}\sum_{l = 1}^{N_{\text{traj}}} \delta(\mathbf{X} - \mathbf{X}_l(t))\delta(\mathbf{P} - \mathbf{P}_l(t))
	 \end{equation}
	 where $N_{\text{traj}}$ denotes the total number of trajectories. At time $t=0$, $\mathbf{X}_l(0)$ and $\mathbf{P}_l(0)$ are sampled according to the  initial Wigner distribution $\rho_{W}(\mathbf{X}, \mathbf{P}, 0)$. Similarly, 	the expectation value of operator $\hat{A}$ can  be evaluated by averaging over trajectories:
	 \begin{equation}\label{eq:A_sample_traj}
	 \avg{\hat{A}} \approx \frac{1}{N_{\text{traj}}}\sum_{l = 1}^{N_{\text{traj}}} A_l(t)
	 \end{equation}
	 where $A_l(t)$ denotes the classical correspondence of $\hat{A}$ for the $l$-th trajectory. 
	 For each classical trajectory $(\mathbf{X}_l(t), \mathbf{P}_l(t))$, the time evolution obeys Hamiltonian mechanics:
	\begin{subequations}\label{eq:dxdpdt_Hamiltonian}
		\begin{align}
		\dot{\mathbf{X}}_l(t) &= \frac{\partial H\left(\mathbf{X}_l, \mathbf{P}_l \right) }{\partial \mathbf{P}_l(t)} \\
		\dot{\mathbf{P}}_l(t) &= -\frac{\partial H\left(\mathbf{X}_l, \mathbf{P}_l \right) }{\partial \mathbf{X}_l(t)}
		\end{align}
	\end{subequations}
	Eqs. \eqref{eq:rhoW_sample_traj}-\eqref{eq:dxdpdt_Hamiltonian} constitute MMST dynamics, i.e., just linearized semiclassical dynamics \red{(LSC) using the MMST mapping to generate the initial value representation (IVR)} so that one can sample some quantum zero-point effects through the Wigner representation.

	Let us now discuss how we will treat a few remaining technical issues in our MMST calculations below: (i) how to correctly sample the initial phase space distribution? (ii) how to efficiently evolve independent trajectories? (iii) how to calculate observables for MMST dynamics?

	\subsubsection{Practical Implementation of MMST Dynamics}
		
	\paragraph{Initial sampling of phase space distribution}
	For $N$ TLSs coupled to a multimode cavity, we assume that at time $t = 0$, the electronic and photonic Wigner distributions are totally decoupled:
	\begin{equation}
	\rho_{W}(\mathbf{X}, \mathbf{P}, 0) = \prod_{\alpha k} \rho^{e}_{W}(x_{\alpha k}, p_{\alpha k}, 0)\otimes\prod_{j}\rho^{p}_{W}(X_j, P_j, 0)
	\end{equation}
	For the photonic DoFs, we assume that $\rho^{p}_{W}(X_j, P_j, 0)$ for each photon mode  obeys the zero-temperature vacuum distribution:
	\begin{equation}\label{eq:Wigner_distr_photon}
	\rho_{W}^p(P_j, X_j, 0) = \frac{1}{\pi}e^{-\frac{P_j^2}{\omega_j} - \omega_j X_j^2}
	\end{equation}
	For the electronic DoFs, \red{however, there is an interesting twist about how to sample the initial distribution. In principle, for TLS $\alpha$, if one starts from the \textit{electronic} ground state, one should initialize $\rho_W^{e}(x_{\alpha e}, p_{\alpha e}, 0)$ from the Wigner distribution for the harmonic oscillator \textit{ground} state, and initialize $\rho_W^{e}(x_{\alpha g}, p_{\alpha g}, 0)$ from the Wigner distribution for the harmonic oscillator \textit{first excited} state, which can be negative. In principle, we avoid this question here by using the simplest square distribution, }
	i.e., we will  first write $x_{\alpha k}$ and $p_{\alpha k}$ in  action-angle coordinates:
	\begin{subequations}\label{eq:xp_to_action_angle}
		\begin{align}
		x_{\alpha k} &= \sqrt{2n_{\alpha k}}\red{\cos\theta_{\alpha k}} \\
		p_{\alpha k} &= \sqrt{2n_{\alpha k}}\red{\sin\theta_{\alpha k}}
		\end{align}
	\end{subequations}
	and then sample the action $n_{\alpha k} =\frac{1}{2}\left(x_{\alpha k}^2 + p_{\alpha k}^2\right)$ and angle $\theta_{\alpha k} = \red{\arctan\left(\frac{p_{\alpha k}}{x_{\alpha k}}\right)}$ by
	\begin{subequations}
		\label{eq:action_angle_distribution}
		\begin{align}
		\label{eq:action_distribution}
		n_{\alpha k} &\in [n_{k}^{(\alpha)}, n_{k}^{(\alpha)}+2\gamma] \\
		\label{eq:angle_distribution}
		\theta_{\alpha k} & \in [0, 2\pi)
		\end{align}
	\end{subequations}
	Eq. \eqref{eq:action_distribution} implies that the action ($n_{\alpha k}$) obeys a uniform distribution in the interval  $[n_{k}^{(\alpha)}, n_{k}^{(\alpha)}+2\gamma]$, where $n_g^{(\alpha)} = 1$ and $n_e^{(\alpha)} = 0$ if the TLS is in the ground state, and $n_g^{(\alpha)} = 0$ and $n_e^{(\alpha)} = 1$ if the TLS is in the excited state. Eq. \eqref{eq:angle_distribution} implies that the angle ($\theta_{\alpha k}$) is a random angle in the interval $\left[0, 2\pi\right)$. After sampling the action-angle coordinates as Eq. \eqref{eq:action_angle_distribution}, we transform $n_{\alpha k}$ and $\theta_{\alpha k}$ to $x_{\alpha k}$ and $p_{\alpha k}$ using Eq. \eqref{eq:xp_to_action_angle}. 	
	Note that in sampling $n_{\alpha k}$, we set $\gamma = 0.45$ rather than $\gamma = 1/2$.

	\red{A few words are now in order regarding the choice of $\gamma$. Strictly speaking, according to the MMST mapping, the correct electronic ZPE should be  $\gamma=1/2$. However, in practice, the situation is far more complicated. First, by now, there is ample evidence that suggest choosing $\gamma$ different  from $1/2$ can yield far better results\cite{Muller1999,Cotton2013}. Second, from a theoretical point of view, we know that propagating independent trajectories can be dangerous and lead to more or less ZPE leakage\cite{Muller1999}. Third, when sampling trajectories, one should in principle sample from a Wigner distribution of the harmonic oscillators, but this distribution can be negative (which can lead to  complications).  As such, it is common nowadays to sample trajectories from phase space in square or other distributions (rather than just a Wigner transform). And moreover, given that one is not going to use a Wigner distribution, one must be all the more tempted to use a different value of $\gamma$ as well.  
	\blue{The rationale would be as follows: in contrast to a truly quantum theory --- where ZPE is an intrinsic property of the photon or electron itself ---  classical mechanics has no such intrinsic property; and therefore, because MMST dynamics treat the ZPE classically, the correct electronic ZPE in MMST should also depend on the correlation between electron and the photonic bath, and a general theory to determine the value of ZPE in MMST dynamics, has not been developed unfortunately.}
	Altogether, this leads to the notion of treating $\gamma$ as an optimization parameter, and there is indeed a large literature discussing how to choose $\gamma$ so as to minimize zero point energy leakage without compromising short time dynamics\cite{Muller1999,Habershon2009}. For coupled electron-nuclear dynamics, Cotton and Miller suggest $\gamma=0.366$ is a good choice\cite{Cotton2013}.} For our part, by investigating the electronic dynamics for a TLS starting from the ground state (see Figs. \ref{fig:ground_state_popu}-\ref{fig:ground_state_phase_space}), we find $\gamma=0.45$ leads to the stability of the electronic ground state exactly. In practice, we also find that changing $\gamma=0.45$ to other values (say, 0.5) does not significantly alter our results below.

	\paragraph{Efficient propagation of independent trajectories: coupled Maxwell-Schr\"odinger equations}
	For a given trajectory, after sampling the initial conditions of $\mathbf{X} = \{x_{\alpha k}, X_j\}$ and $\mathbf{P} = \{p_{\alpha k}, P_j\}$ as in Eqs. \eqref{eq:Wigner_distr_photon}-\eqref{eq:action_angle_distribution}, one can directly propagate $\mathbf{X}$ and $\mathbf{P}$ using the Hamiltonian equations of motion (Eq. \eqref{eq:dxdpdt_Hamiltonian}). However, for the coupled electron-photonic system, directly propagating Eq. \eqref{eq:dxdpdt_Hamiltonian} is \red{more computationally expensive than necessary; after all,} in the coupling term of the classical Hamiltonian ($h_{kl}$ term in Eq. \eqref{eq:H_Wigner_MMST}), each TLS is coupled to all photonic modes through $h_{kl} = \sum\limits_{j=1}^{M}\red{\hbar}
	g_j^{(\alpha)}\sqrt{\frac{\omega_{j}}{2\hbar}} X_j$, which \red{implies that for any TLS $\alpha$, propagating $x_{\alpha k}$ and $p_{\alpha k}$ requires a summation over all photon modes  at each time step. This additional summation  loop  becomes computational more and more expensive when the number of TLSs is large.}
	To avoid this summation, \red{instead of propagating $X_j$ and $P_j$, one can propagate the collective variables $\vE(r, t)$ and $\vB(r, t)$, i.e., the}
	 classical EM fields:
	\begin{subequations}\label{eq:XP_2EB_classical}
		\begin{align}
		\label{eq:E_def}
			\vE(r, t) &= \sum_{j=1}^{M}  \sqrt{\frac{2}{\epsilon_0 L}}\omega_{j} X_j(t)\sin(k_j r) \ve_E\\
			\vB(r, t) &= \sum_{j=1}^{M} \sqrt{\frac{2\mu_0}{L}}P_j(t) \cos(k_j r) \ve_B
		\end{align}
	\end{subequations}
	where $\ve_E$ ($\ve_B$) denotes the the direction of the electric (magnetic) field.

	\red{For the electronic DoFs, since we regard the photonic degrees of freedom as classical objects,  propagation of the classical Hamiltonian equations for the electronic ($x_{\alpha k}$, $p_{\alpha k}$) degrees of freedom  (Eq. \eqref{eq:dxdpdt_Hamiltonian})  can be replaced exactly by propagation of the complex variables $c_{\alpha k} $, which is defined as
	\begin{equation}\label{eq:action-angle-to-wavefunction}
	c_{\alpha k} \red{ = \frac{1}{\sqrt{2}}\left(x_{\alpha k} + i p_{\alpha k}\right)} = \sqrt{n_{\alpha k}}e^{i\theta_{\alpha k}}.
	\end{equation}
	Thus, effectively, one can propagate
	 the electronic wave function ($\ket{\psi_{\alpha}}$) for a TLS 
	\begin{equation}
	\red{\ket{\psi_{\alpha}}} = c_{\alpha g}\ket{\alpha g} +c_{\alpha e}\ket{\alpha e}
	\end{equation}
	 instead of $\{x_{\alpha g}, x_{\alpha e}, p_{\alpha g}, p_{\alpha e}\}$.  More generally, it is worth emphasizing that propagating a quantum Schr\"odinger equation (see Eq. \eqref{eq:Schrodinger} below) with a one-body Hamiltonian in an $N$ dimensional vector space is entirely equivalent to propagating classical mechanics with $N$ harmonic oscillators --- which is the very essence of the MMST mapping.  In practice, from our perspective, we have chosen to propagate the quantum electronic Schr\"odinger equation here  (rather than classical Hamiltonian equations of motion) so that the reader can more easily compare and digest the present work with existing work in the area of semiclassical electrodynamics\cite{Crisp1969,Lorin2007,Sukharev2011,Gao2012,Li2016,Li2018Spontaneous}.}

	 To that end, for each trajectory, one can equivalently solve the coupled Maxwell-Schr\"odinger equations:
	\begin{subequations}\label{eq:coupled_maxwell_schrodinger}
			\begin{align}\label{eq:Maxwell}
		\frac{\partial}{\partial t} \vB(r, t) &= - \vnabla \times \vE(r, t) \\
		\frac{\partial}{\partial t} \vE(r, t) &= c^2 \vnabla \times \vB(r, t) - \frac{\vJ(r, t)}{\epsilon_0},
		\end{align}
		\begin{align}\label{eq:Schrodinger}
		i\hbar \frac{d}{dt} \ket{\psi_\alpha(t)} =  \left[\frac{1}{2}\hbar\omega_{0}\hsigma_z^{(\alpha)} - \int dr \vE_{\perp}(r)\cdot \hP^{(\alpha)}(r)\right] \ket{\psi_\alpha(t)}
		\end{align}
	\end{subequations}
	Here, the classical current density ($\vJ(r, t)$) is calculated in a mean-field way:
	\begin{equation}
	\label{eq:J-Ehrenfest}
	\begin{aligned}
	\vJ(r, t) = \sum\limits_{\alpha =1}^{N}\frac{\partial }{\partial t}\tr{\hrho^{(\alpha)}(t)\hP^{(\alpha)}(r)}
	\end{aligned}
	\end{equation}
	where $\hrho^{(\alpha)}(t) = \ket{\psi_\alpha(t)}\bra{\psi_\alpha(t)}$. The atomic polarization density operator $\hP^{(\alpha)}(r)$ is defined as
	\begin{align}\label{eq:hP}
	\hP^{(\alpha)}(r) = \mu_{ge}^{(\alpha)}\he_d^{(\alpha)}\delta(r - r_\alpha)\left(\hsigma_{+}^{(\alpha)} + \hsigma_{-}^{(\alpha)}\right)
	\end{align}	
	where $\he_d^{(\alpha)}$ denotes the unit vector along the polarization. In 1D, we can simply assume that $\he_d^{(\alpha)}$ is parallel to the direction of the \textit{E} field. Note that in Eq. \eqref{eq:Schrodinger} for 1D problems, one can simply replace $\int dr \vE_{\perp}(r)\cdot \hP^{(\alpha)}(r)$ by $\int dr \vE(r)\cdot \hP^{(\alpha)}(r)$ in numerical calculations and ignore the distinction between \red{the} total and transverse \textit{E} field. 
	\red{Note that, by applying Eqs. \eqref{eq:gj}, \eqref{eq:E_def} and \eqref{eq:hP}, $\int dr \vE(r)\cdot \hP^{(\alpha)}(r) = \sum_{j = 1}^{M}\sqrt{2\hbar\omega_{j}}g_j^{(\alpha)}X_j(t)\left(\hsigma_{+}^{(\alpha)} + \hsigma_{-}^{(\alpha)}\right)$. Compared with the quantum light-matter coupling term (Eq. \eqref{eq:H_tot_QED_1})  $\sum_{j=1}^{M}\hbar g_j^{(\alpha)}\left(\hat{a}^{\dagger}_j + \hat{a}_j\right)\left(\hsigma_{+}^{(\alpha)} + \hsigma_{-}^{(\alpha)}\right) = \sum_{j=1}^{M} g_j^{(\alpha)}\sqrt{{2\hbar \omega_{j}}}  \hat{X}_j\left(\hsigma_{+}^{(\alpha)} + \hsigma_{-}^{(\alpha)}\right)$, one can ascertain the equivalence of these two quantities by assuming $\hat{X}_j$ is classical ($X_j(t)$).}

	In the end, to efficiently evolve the MMST dynamics for the coupled electron-photonic dynamics, one needs to first sample the initial distribution of $X_j$ and $P_j$ according to Eq. \eqref{eq:Wigner_distr_photon} and second transform \red{all} initial coordinates into EM fields according to Eq. \eqref{eq:XP_2EB_classical}. Similarly, for the electronic DoFs, after sampling the initial condition using the action-angle coordinates (Eq. \eqref{eq:action_angle_distribution}), one transforms coordinates to \red{the  wave function picture using} Eq. \eqref{eq:action-angle-to-wavefunction}. Finally, for each trajectory, one evolves the coupled Maxwell-Schr\"odinger equations \red{in} Eq. \eqref{eq:coupled_maxwell_schrodinger}.
	
	\paragraph{Calculating observables}
	Finally, within MMST dynamics, because  we explicitly take both the electronic and photonic ZPEs into account, when calculating observables, we need to exclude all ZPE effects. 
	
	\red{For example, because we initialize the electronic DoFs by a square distribution (see Eqs. \eqref{eq:xp_to_action_angle} and \eqref{eq:action_angle_distribution}), after transforming to the wave function representation (using $c_{\alpha k}  = \frac{1}{\sqrt{2}}\left(x_{\alpha k} + i p_{\alpha k}\right) $),  this implies that our initial wave function (for each TLS $\alpha$) will on average be normalized  as $|c_{\alpha g}|^2 + |c_{\alpha e}|^2= 1 + 2\gamma$, i.e., on average each electronic level contains an additional $\gamma$. We call this extra normalization electronic ZPE. And in practice, this ZPE is essential for recovering many electrodynamical phenomena, e.g. spontaneous emission. See Sec. \ref{sec:results}. Nevertheless, when calculating the electronic population for the excited state, one must subtract the electronic ZPE from the raw average to obtain a reasonable answer:}
	\begin{equation}\label{eq:rhoee_MMST}
	\rho_{ee}^{(\alpha)}(t) = \avg{|c_{\alpha e}(t)|^2}_l - \gamma
	\end{equation}
	where $\avg{\cdots}_l$ denotes taking the ensemble average over trajectories.  Of course, if we were to sample only the photonic ZPE and not  the electronic ZPE, we could calculate the excited state population by  	$\rho_{ee}^{(\alpha)}(t) = \avg{|c_{\alpha e}(t)|^2}_l$. Note that Cotton and Miller have used more sophisticated approaches (i.e., using symmetric windowing functions) with strong results for  evaluating  electronic populations\cite{Cotton2013,Cotton2016,Cotton2019}, but we find that using Eq. \eqref{eq:rhoee_MMST} is good enough for our simulations below.

	Similarly, when calculating the E-field intensity, the correct expression must be
	\begin{equation}\label{eq:I_MMST}
	I(r, t) = \epsilon_0 \avg{E_z^2(r, t)}_l - \epsilon_0\sum_j \varepsilon_j^2\sin^2(k_j r)
	\end{equation}
	Again, if we were to sample only the electronic ZPE and  not  the photonic ZPE, we would  calculate the intensity to be  $I(r, t) = \epsilon_0 \avg{E_z^2(r, t)}_l$.
	\red{Finally, note that	 when calculating observables, Eq. \eqref{eq:rhoee_MMST} may predict negative populations or negative forces due to the subtraction of ZPEs\cite{Shi2004,Bonella2001,Bonella2005,Kelly2012,Bellonzi2016}. So far, in our calculations  with harmonic photonic fields coupled linearly to electronic DoFs, we have found empirically that such side effects can be minimized by optimizing the value of electronic ZPE ($\gamma$).}

	\subsubsection{Comments on MMST dynamics}
	\label{sec:MMST_comments}

	\red{
		After reviewing  MMST dynamics, we can summarize  the relationship between MMST dynamics and other approaches in Fig. \ref{fig:demo}, where (from left to right) different methods are arranged in  descending order of computational cost.  Working from the full QED Hamiltonian, one can propagate the light-matter dynamics by  working under a truncated Hilbert space, e.g., by applying a CIS or CISD (where "D" denotes the doubly excited states) approximation, or by applying  Dicke's model\cite{Dicke1954,Gross1982}, i.e.,  only symmetric wave functions are considered. By contrast, as we showed above, one can also  transform the QED Hamiltonian to Wigner phase space dynamics through an MMST mapping. Exact Wigner dynamics will yield exact quantum dynamics; however, one usually performs MMST dynamics by assuming independent trajectories (which leads to enormous computational savings). If the ZPEs are further neglected, MMST dynamics can be reduced to the conventional Maxwell-Schr\"odinger equations, where only a single trajectory is needed for a given simulation.
		
			\begin{figure}
	\includegraphics[width=1.0\linewidth]{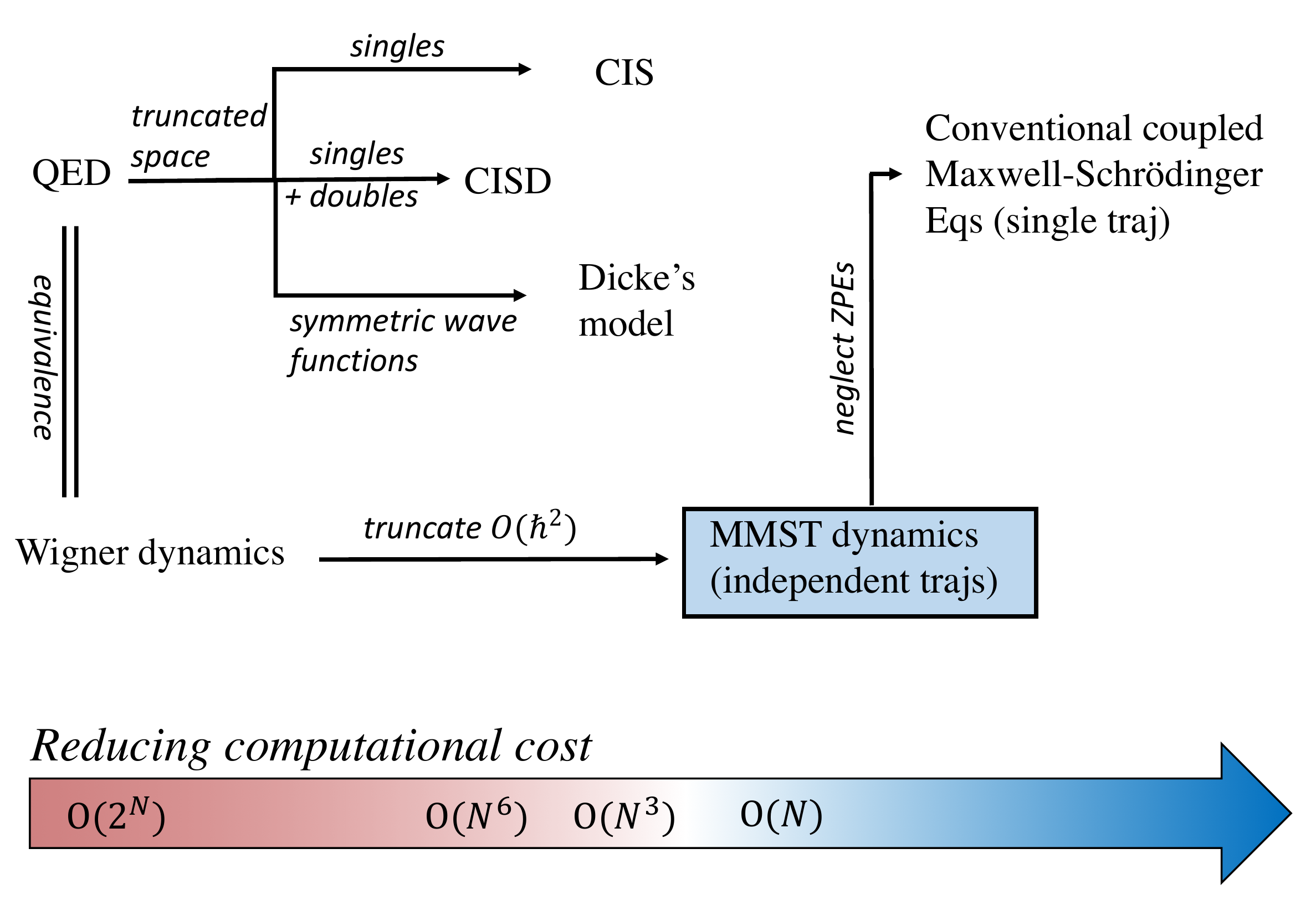}
	\caption{Relationship between MMST dynamics and other quantum and semiclassical approaches. From left to right, approaches are arranged in the descending order of  computational cost. }
	\label{fig:demo}
\end{figure}
	
	\blue{Note that MMST dynamics do not invoke a rotating-wave approximation (unlike the CIS approximation). Instead, for MMST} dynamics, 
	the major approximation is to propagate \textit{independent} trajectories, which originates from truncating all $O(\hbar^2)$ terms in Eq. \eqref{eq:drhodt_Wigner_MMST}\cite{footnote3}. According to the Wigner expansion, the leading term dropped is $- \frac{\hbar^2}{24}\frac{\partial^3 H}{\partial \mathbf{X}^3}\frac{\partial^3 \rho_{W}}{\partial \mathbf{P}^3}$, and given the form of $H$ in Eq. \eqref{eq:H_Wigner_MMST}, whereby the only  cubic term is proportional to the light-matter coupling strength $g_j^{(\alpha)}$, it is clear that
	(i) MMST dynamics will be  less accurate when the light-matter coupling strength $g_j^{(\alpha)}$ is very, very large; (i) of course, MMST dynamics become less accurate for longer and longer times;  By contrast, if the light-matter coupling is not very large ($g_j^{(\alpha)} \ll \omega_{0}, \omega_{j}$) and  the simulation time is not very long, MMST dynamics should predict promising results, which will be shown in Sec. \ref{sec:results}. 
	}

	\section{Simulation Details}\label{sec:simulation_details}

		\subsection{Quantum simulation}
		
		For our reference quantum simulations below, we will largely adapt the parameters in Ref. \cite{Buzek1999}. Using natural units ($[c] = [\hbar] = [\epsilon_0]  = 1$), we consider the case that each TLS has the same energy ($\omega_{0} = 100$) and the same transition dipole moment ($\mu_{ge} = \sqrt{\pi} / 10$), and the cavity length to be $L = 2\pi$. Note that for the field Hamiltonian, a hard cutoff  for the maximum photon energy is taken such that $\omega_{j}^{\text{max}} = 2\omega_{0} = 200$. In total, $M = 400$ photon modes are used,  so that neighboring photon modes have energy difference ($\Delta \omega$) as $0.5$.  
		\red{For a TLS ($\alpha$) located at the middle of the cavity ($r_{\alpha} = L/2 = \pi$), given the above parameters, according to Eq. \eqref{eq:gj}, the light-matter coupling strength with mode $j$ is $g_j^{(\alpha)} = \sqrt{\frac{\omega_j}{\hbar\epsilon_0 L}}\mu_{ge}^{(\alpha)}\sin(k_j r_\alpha) =  \frac{\sqrt{j}}{20} \sin(\frac{j\pi}{2})$.}
		After constructing the full quantum Hamiltonian in Eq. \eqref{eq:H_tot_QED} using the CIS basis, the wave function at any given time is directly evaluated by calculating $\ket{\Psi_{\text{CIS}}(t)} = \exp\left(-\frac{i}{\hbar}\hH t\right)\ket{\Psi_{\text{CIS}}(0)}$. We calculate expectation values using Eqs. \eqref{eq:population_operator}-\eqref{eq:intensity_operator}.

		\subsection{MMST simulation}
		For MMST dynamics, we propagate Maxwell equations using the finite-difference time-domain (FDTD) technique\cite{Taflove2005}, according to which the \textit{E} field and \textit{B} field are propagated in a staggered grid, as suggested by Yee\cite{Yee1966}. In a 1D cavity, if we assume that the \textit{E} field orients along \red{the} $z$-axis and \red{the} B-field orients along the $y$-axis, according to \red{the}  FDTD technique,  Maxwell equations can be numerically discretized as:
		\begin{widetext}
	\begin{subequations}
		\label{eq:FDTD_algorithm}
		\begin{align}
		\label{eq:FDTD_algorithm_Dz}
		E_z^{m + \frac{1}{2}}(k) &= E_z^{m - \frac{1}{2}}(k) + \frac{\Delta t}{\Delta x}\left[B_y^m\left(k + \frac{1}{2}\right) - B_y^m\left(k - \frac{1}{2}\right)\right] \\
		\label{eq:FDTD_algorithm_Ez}
		E_z^{m+ \frac{1}{2}}(k) &= \frac{1}{\epsilon_0}\left[E_z^{m + \frac{1}{2}}(k) - P_z^{m + \frac{1}{2}}(k)\right] \\
		\label{eq:FDTD_algorithm_boundary}
		E_z^{m + \frac{1}{2}}(0) &= E_z^{m + \frac{1}{2}}(N_{\text{grids}} - 1) = 0\\
		\label{eq:FDTD_algorithm_By}
		B_y^{m + 1}(k+ \frac{1}{2}) &= B_y^{m}(k+\frac{1}{2}) - \frac{\Delta t}{\Delta x}\left[E_z^{m + \frac{1}{2}}\left(k +1\right) - E_z^{m - \frac{1}{2}}\left(k \right)\right] 
		\end{align}
	\end{subequations}
	\end{widetext}
	where $m = 0, 1,  \cdots$ denotes the index for time step,  and $k = 0, 1, \cdots, N_{\text{grids}}-1$ denotes the index for the 1D spatial grids, and $\Delta x$ denotes the grid spacing. Here, in Eq. \eqref{eq:FDTD_algorithm_Ez}, $P_z(r, t) = J_z(r, t) \Delta t$, and $J_z(r, t) = \sum\limits_{\alpha =1}^{N} -2\omega_{0}\text{Im}\left[\rho^{(\alpha)}_{ge}(t)\right]\xi_\alpha(r)$, where $\xi_\alpha(r)$ denotes the spatial distribution of the $i$-th atomic polarization density. To represent a collection of TLSs, we would like to follow Eq. \eqref{eq:hP} and set $\xi_\alpha(r) = \mu_{ge}\delta(r - r_\alpha)$. However, in practice, for numerical stability, we use a Gaussian form instead: $\xi_\alpha(r) = \mu_{ge}\sqrt{2\pi}\sigma e^{-\frac{(r - r_\alpha)^2}{2\sigma^2}}$, where $\sigma$ denotes the width of each TLS.  Eq. \eqref{eq:FDTD_algorithm_boundary} defines the boundary conditions for the cavity: the E-field values at the boundaries are forced to be  exactly zero. As far as other parameters are considered, we set $N_{\text{grids}} = 5001$ grid points with $\Delta x = L / (N_{\text{grids}} - 1) = 2\pi / 5000$, $\Delta t = \Delta x /2c$, and $\sigma = 10^{-3}$.
	
	Within FDTD, because the E- and B-fields are propagated in a staggered space-time grid (Yee cell),  the initial values of the E- and B-fields should also be set with a Yee cell framework:
		\begin{subequations}
		\begin{align}
		E_z^{0 + \frac{1}{2}}(k) &= \sum_{j} \sqrt{\frac{2}{\epsilon_0 L}}\omega_{j} X_j^{0+\frac{1}{2}} \sin(k_j r_k) \\
		B_y^{0 + 1}\left(k + \frac{1}{2}\right) &= \sum_{j} \sqrt{\frac{2\mu_0}{L}} P_j^{0 + 1}\cos\left[k_j \left(r_k + \frac{\Delta x}{2}\right)\right]
		\end{align}
		where $r_k = k\Delta x$.
		Note that  after sampling $P_j^{0 + \frac{1}{2}}$ and $X_j^{0 + \frac{1}{2}}$ according to Eq. \eqref{eq:Wigner_distr_photon} at the same time ($0 + \frac{1}{2}\Delta t$), we need to  evolve $P_j$ for another half time step $P_j^{0 + 1} = P_j^{0 + \frac{1}{2}} - \omega_j^2 X_j^{0 + \frac{1}{2}}\frac{\Delta t}{2}$ to calculate $B_y^{0 + 1}$.
	\end{subequations}
	
	To be compatible with FDTD, we propagate the electronic wave function via the split operator technique (which also uses half time steps):
	\begin{equation}
	\label{eq:split-operator}
	\ket{\red{\psi_{\alpha}}^{m+ \frac{1}{2}}} = e^{-\frac{i}{\hbar}\hat{V}^{(\alpha)} \frac{\Delta t}{2}}e^{-\frac{i}{\hbar}\hat{H}_{\text{s}}^{(\alpha)} \Delta t}e^{-\frac{i}{\hbar}\hat{V}^{(\alpha)}  \frac{\Delta t}{2}} \red{\ket{\psi_{\alpha}}^{m - \frac{1}{2}}}
	\end{equation}
	Here, $\hH_s^{(\alpha)} = \frac{1}{2}\hbar\omega_{0}\hsigma_z^{(\alpha)}$,  $\hV^{(\alpha)} = - \int dr E_z(r) \hat{P}_z^{(\alpha)}(r)$, and $\hat{P}^{(\alpha)}_z = \mu_{ge} \xi_\alpha(r)\left(\hsigma_{+}^{(\alpha)} + \hsigma_{-}^{(\alpha)}\right)$. The initial $\ket{\red{\psi_{\alpha}}^{m + \frac{1}{2}}}$ is sampled according to Eqs. \eqref{eq:action_angle_distribution} and \eqref{eq:action-angle-to-wavefunction}. 
	Eqs. \eqref{eq:FDTD_algorithm} and \eqref{eq:split-operator} form the split-operator-finite-difference-time-domain (SO-FDTD) propagator. 

	During an MMST simulation,  we calculate observables with Eqs. \eqref{eq:rhoee_MMST} and \eqref{eq:I_MMST} by averaging over a swarm of trajectories.
	Unfortunately, for our simulations, in order to eliminate random noise from sampling ZPEs (for both electrons and photons), many trajectories are needed. In order to calculate smooth population dynamics, we require $10^4$ trajectories. In order to obtain a smooth distribution of the E-field intensity, we need $10^6$ trajectories because we sample $400$ photon modes.
	
	\red{The simulation code (written in C++) and raw data are available at Github\cite{Li2019Github}.}

	\section{Results}\label{sec:results}
	
	MMST dynamics provide a systematic way  to evaluate quantum dynamics for a coupled electron-photonic system. In this section, by modeling three fundamental cQED processes, we will show that MMST dynamics sometimes recover very accurate quantum dynamics. The three fundamental processes will be as follows:
	 (i) spontaneous emission for a TLS in a cavity, (ii) modification of the spontaneous emission rate  for a chain of TLSs in a cavity when only the middle TLS is excited initially, or (iii) Dicke's superradiance and subradiance when all TLSs are \red{initially excited}  in a cavity. Before addressing these processes, however, we will make a short digression  to show how MMST balances both electronic and photonic ZPE effects to correctly achieve the stability of the electronic ground state, which connects to notions of  radiative self-interaction and vacuum fluctuations suggested long ago\cite{Dalibard1982,Dalibard1984}.

	\subsection{Stability of Electronic Ground State and the Physical Meanings of Sampling}
	
	In classical electrodynamics, the electronic ground state is never stable. For example, for a classical model of  a hydrogen atom (i.e., a negatively charged electron oscillating around a positively charged nucleus with a certain orbital), the oscillating electron keeps loosing energy due to the radiative self-interaction until the electron collapses to the nucleus. However, according to Dalibard, Dupont-Roc, and Cohen-Tannoudji's (DDC's) QED calculation\cite{Dalibard1982,Dalibard1984}, the electronic ground state is stable because the energy gain from vacuum fluctuations  exactly balances the energy loss due to radiative self-interaction.   In their interpretation,  vacuum fluctuations denote how "\textit{the reservoir fluctuates and interacts with the polarization induced in the small system}", and self-interaction denotes how "\textit{the small system fluctuates and polarizes the reservoir which reacts back on the small system}"\cite{Dalibard1982}\red{\cite{footnote2}}.

	\begin{figure}
	\includegraphics[width=1.0\linewidth]{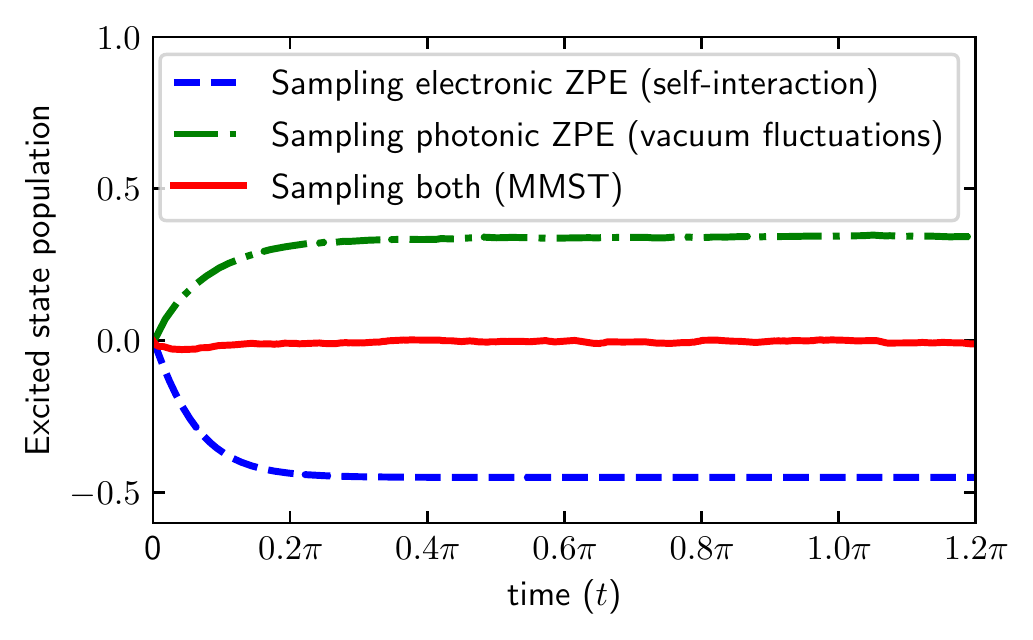}
	\caption{Time evolution of the excited state population for a TLS in a 1D cavity starting from electronic ground state.
	We propagate MMST dynamics by (i) sampling both electronic and photonic ZPEs (red \red{solid}  line), (ii) sampling only electronic ZPE (blue \red{dashed}  line), and (iii) sampling only photonic ZPE (green \red{dash-dot}  line). Note that, consistent with the DDC interpretation of QED, we identify the effect of sampling electronic (photonic) ZPE as corresponding to radiative self-interaction (vacuum fluctuations): self-interaction leads to the breakdown of electronic ground state while vacuum fluctuations lead to spontaneous absorption. By contrast, considering both effects leads to the stability of the electronic ground state.
	For parameters, the TLS is located at the center of the cavity and all other parameters are the same as Sec. \ref{sec:simulation_details}. 12800 trajectories are averaged for all cases.
	}
	\label{fig:ground_state_popu}
	\end{figure}

	This quantum interpretation can be matched up exactly with the sampling  in MMST dynamics:
	the sampling of photonic ZPE should represent the effect of vacuum fluctuations, and the sampling of electronic ZPE should represent the effect of self-interaction. Thereafter, one can test the DDC interpretation of QED semiclassically  by modeling a single TLS in the ground state coupled to a multimode cavity.

	In Fig. \ref{fig:ground_state_popu}, we show that, for a TLS in the ground state initially, when both electronic and photonic DoFs are sampled with MMST dynamics (red \red{solid}  line),  the electronic excited state population is always almost zero as a function of time, i.e., the electronic ground state is stable. By contrast, sampling only the photonic ZPE (green \red{dash-dot}  line) leads to  spontaneous absorption: the excited state population goes up due to the presence of photonic ZPE, which agrees with the DDC view of vacuum fluctuations \red{--- considering vacuum fluctuations only  leads to  spontaneous absorption.}  Vice versa,   sampling only the electronic ZPE (blue \red{dashed}  line) leads to the breakdown of electronic ground state: the excited state population goes down to negative values, which agrees with the DDC view of self-interaction \red{--- considering self-interaction only  leads to the instability  of electronic ground state.}
	Thus, MMST dynamics are in physical agreement with the DDC interpretation of QED, which adds a further twist to the Miller-Milonni disagreement about semiclassical electrodynamics\cite{Miller1978,milonni1980comments}.
	
		\begin{figure}
		\includegraphics[width=1.0\linewidth]{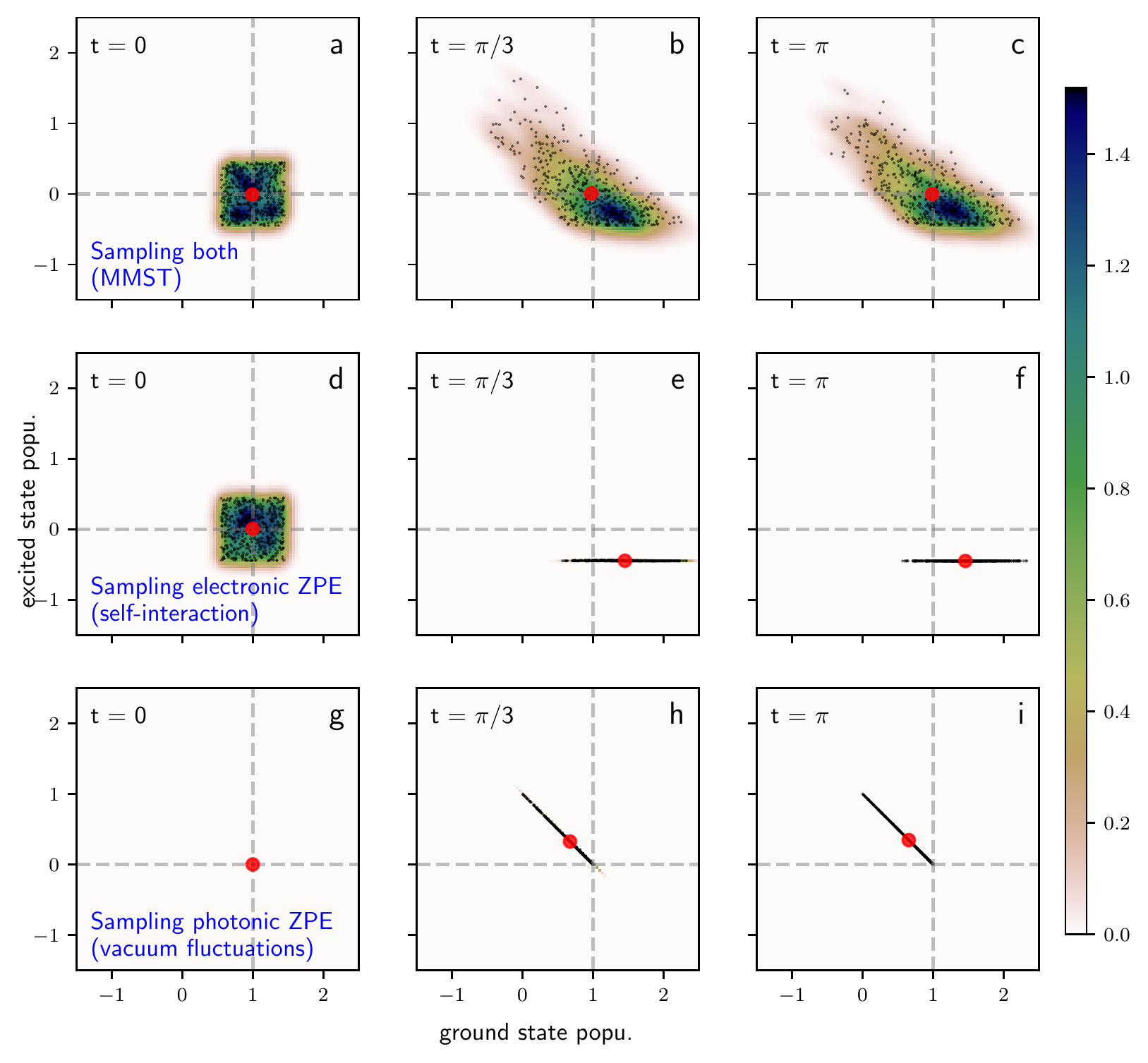}
		\caption{Time evolution of  the phase space distribution for the electronic ground ($x$-axis) and excited state ($y$-axis) populations associated with Fig. \ref{fig:ground_state_popu}. 
		In each subplot, there are 512 black dots, each represents one MMST trajectory. The red dot denotes the average of the trajectories, and the colored contour denotes the electronic phase space distribution calculated by a Gaussian fit of the density of trajectories. 
		From left to right, we plot the electronic phase space distribution at times $t = 0$, $\pi/3$ and $\pi$ when (i) both electronic and photonic ZPEs are sampled (Figs. \ref{fig:ground_state_phase_space}a-c), (ii) only electronic ZPE is sampled (Figs. \ref{fig:ground_state_phase_space}d-f), and (iii) only photonic ZPE is sampled (Figs. \ref{fig:ground_state_phase_space}g-i).
		Note that when both electronic and photonic ZPEs are sampled, the  shape of phase space distribution can be roughly preserved and stabilized in a triangular shape, while sampling either electronic or photonic ZPE leads to the dramatic changes of  shape for the phase space distribution. In practice, sampling only electronic ZPE  leads to negative excited state population, while sampling only photonic ZPE leads to  positive excited state population.}
		\label{fig:ground_state_phase_space}
	\end{figure}

	 Even more interestingly, the phase space evolution of electronic populations can also be evaluated by MMST dynamics.
	 For a TLS starting in the ground state (the same scenario as Fig. \ref{fig:ground_state_popu}),
	 Fig. \ref{fig:ground_state_phase_space} plots the phase space distribution for the ground and excited state populations at different times by sampling trajectories.  In each subplot, one black dot denotes a single trajectory, the red dot denotes the center of all of the trajectories, and the color contour denotes the phase space distribution of electronic populations, which is calculated by a Gaussian fit for the density of trajectories. 
	 
	 When both the electronic and photonic ZPEs are sampled, Figs. \ref{fig:ground_state_phase_space}a-c plot the phase space distribution at time $t = 0$, $\pi/3$ and $\pi$. Similar to Fig. \ref{fig:ground_state_popu}, 
	 the averaged excited state population (red dot) is always zero, showing the stability of the electronic ground state. 
	 Interestingly, the exact shape of the distribution slightly varies from an initial square to a triangle (from $t = 0$ to $\pi/3$) and the shape then stabilizes as a triangle (from $t = \pi/3$ to $\pi$). This finding suggests that initializing with a triangular phase space distribution (at $t = 0$) could be a better choice than a naive square distribution (as in Eq. \eqref{eq:action_angle_distribution}), because the triangular phase space distribution seems to be time-invariant under equilibrium when the TLS interacts with the vacuum field. 
	 Indeed, a recent improvement of MMST dynamics ---  the symmetrical quasiclassical windowing (SQC) dynamics\cite{Cotton2013,Cotton2016} --- suggests that an initial triangular distribution (rather than square) can really improve the performance of SQC dynamics in practice. Nevertheless, because an initial square distribution can already qualitatively predict the preservation of phase space area and can also predict relatively accurate quantum dynamics (which will been shown below), we have stuck with a square distribution  for the electronic DoFs  throughout this paper.
	 
	 By contrast, when only the electronic ZPE is sampled (Figs. \ref{fig:ground_state_phase_space}d-f) or only the photonic ZPE is sampled (Figs. \ref{fig:ground_state_phase_space}g-i), under time evolution, the phase space area of the electronic populations is not preserved even qualitatively, again showing the necessity of sampling both the electronic and photonic ZPEs.

			\begin{figure}
		\includegraphics[width=1.0\linewidth]{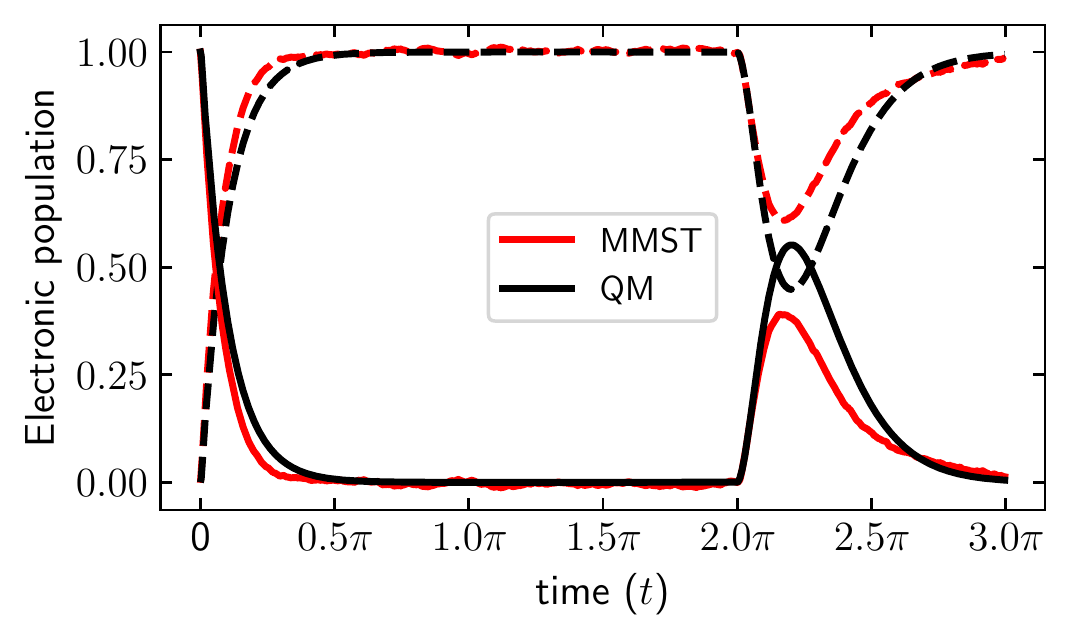}
		\caption{Time evolution of the electronic population for a TLS in a 1D cavity when the TLS starts from the electronic excited state. All other parameters are the same as Fig. \ref{fig:ground_state_popu}. The solid (dashed) lines denote the excited (ground) state population. Note that compared with the QED calculation (black line), MMST (red line) describes both the initial exponential decay and the Poincar\'e recurrence very well.}
		\label{fig:SE_popu_MMST}
	\end{figure}
	
	\begin{figure}
		\includegraphics[width=1.0\linewidth]{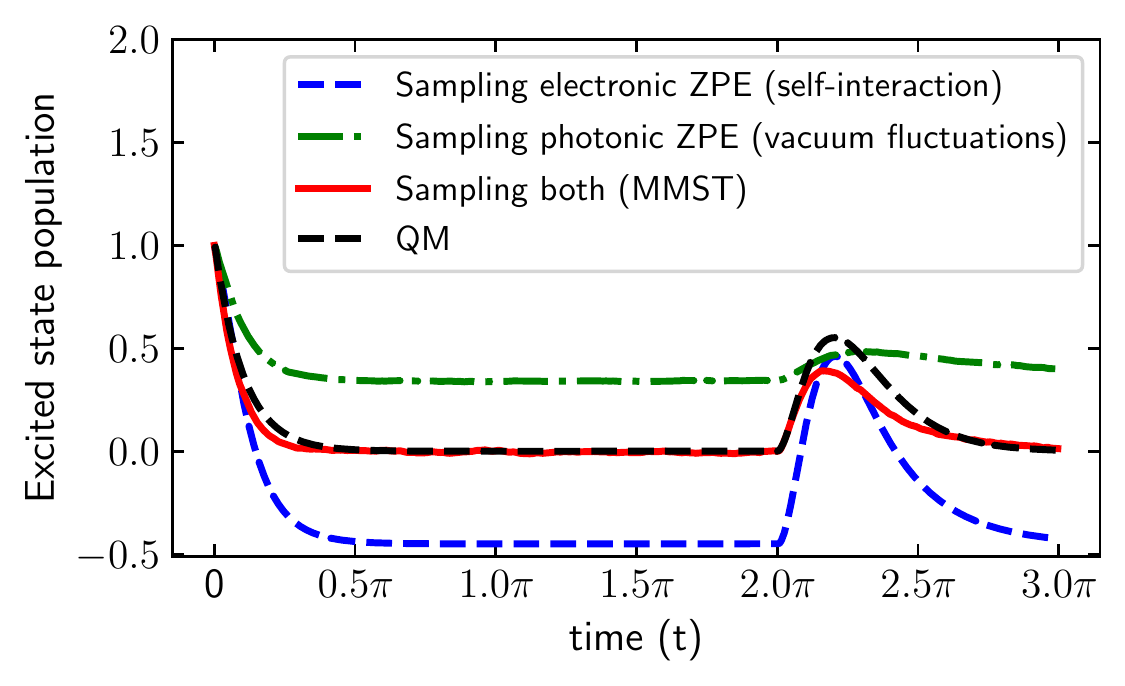}
		\caption{Contributions of self-interaction and vacuum fluctuations for Fig. \ref{fig:SE_popu_MMST} \red{as predicted} by MMST dynamics. Note that at the initial stages ($t \approx 0$), self-interaction and vacuum fluctuations contribute to exponential decay almost equally. By contrast, \red{at} later times ($t = 2\pi$), self-interaction \red{is mostly responsible  for the}  Poincar\'e recurrence.}
		\label{fig:SE_popu_turnoff_samplings}
	\end{figure}

			\begin{figure}
		\includegraphics[width=0.8\linewidth]{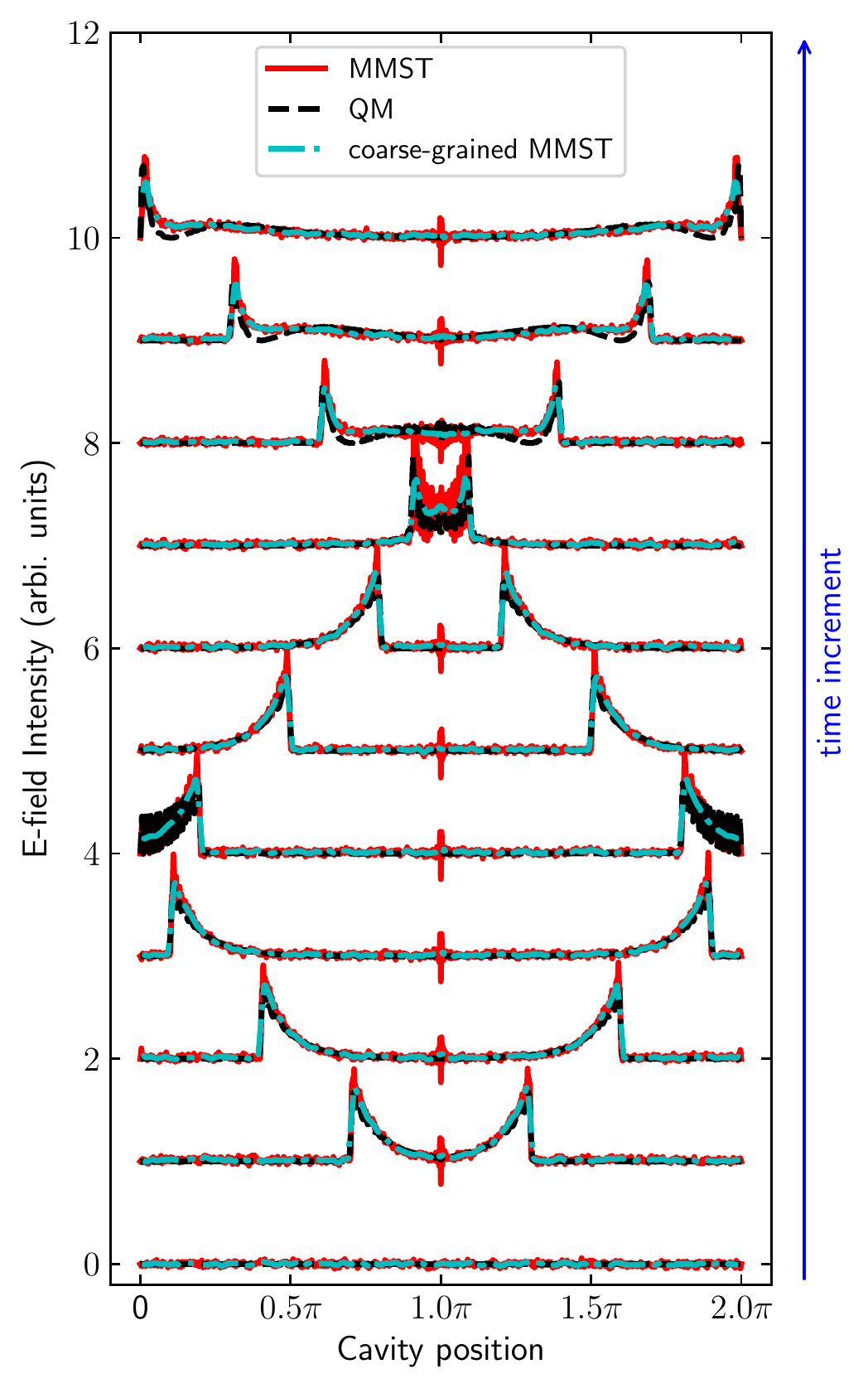}
		\caption{Real-space distribution of the E-field intensity from time $0$ to $3\pi$ (from bottom to top) associated with Fig. \ref{fig:SE_popu_MMST}. Here we average over $1.28 \times 10^6$ trajectories, and all other parameters are the same as in Fig. \ref{fig:SE_popu_MMST}.  Note that MMST dynamics (red \red{solid}  line) are quite accurate here.  There is one hiccup: MMST predicts non-vanishing intensity oscillations (ranging from \textit{negative} to \textit{positive} values) at $r = L/2 = \pi$.  Averaging over the neighboring 50 grid points (\red{cyan dash-dot} line) eliminates such oscillations and the final coarse-grained MMST results agree very well with the QED calculation (black \red{dashed}  line); see Appendix \ref{Appendix: middle_peak} for a detailed discussion for the "middle oscillations".}
		\label{fig:SE_intensity_MMST}
	\end{figure}
	
		\begin{figure*}
		\includegraphics[width=0.7\linewidth]{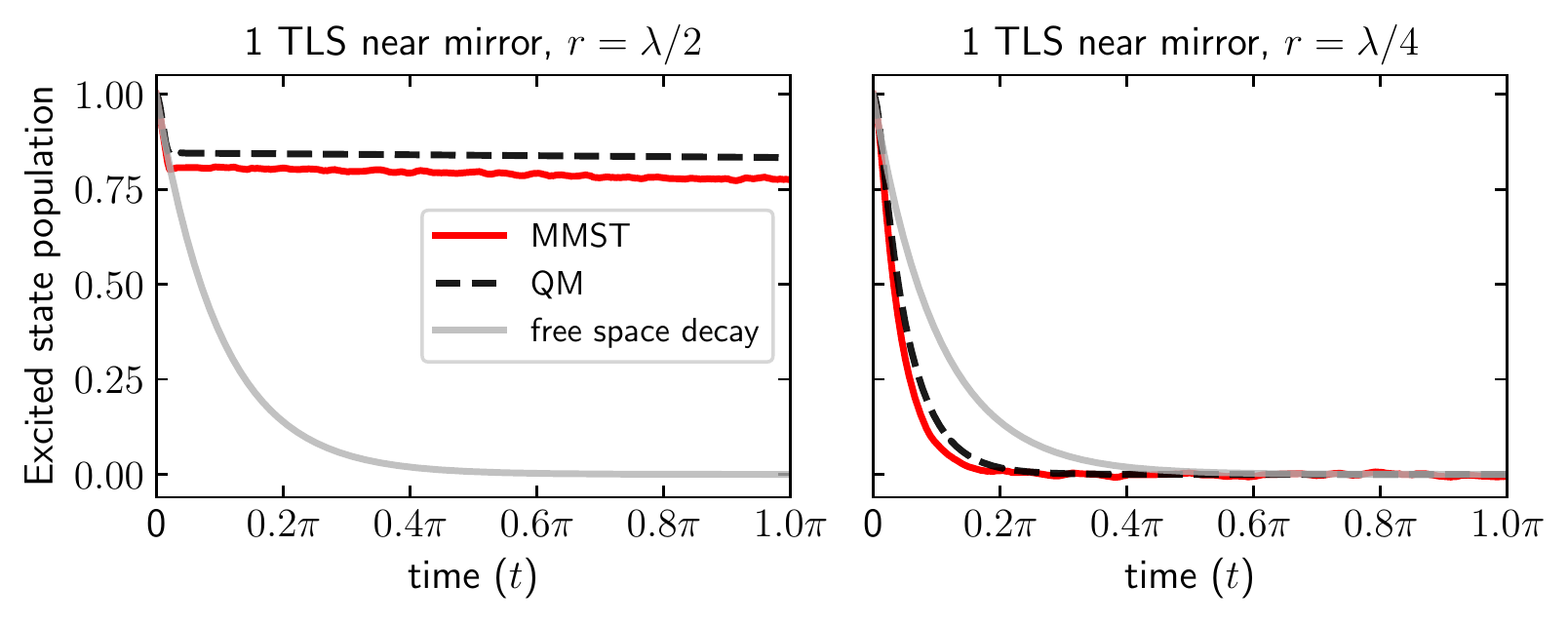}
		\caption{Time evolution of the excited state population when the TLS near the cavity mirror. In the left (right) subplot, the TLS is close to the cavity mirror with distance $r = \lambda/2$ ($r = \lambda/4$), where $\lambda = 2\pi/\omega_{0}$. All other parameters are the same as in Fig. \ref{fig:SE_popu_MMST}. Compared with the free space decay for a single TLS (gray \red{solid}  line), MMST dynamics (red \red{solid}  line) predict the inhibition  (left) or enhancement (right) of spontaneous emission rate when $r = \lambda/2$ or $r = \lambda/4$ respectively; these results agree relatively well with the exact QED calculations (black \red{dashed}   line).}
		\label{fig:SE_near_mirror}
	\end{figure*}
	
	\subsection{Spontaneous Emission for a TLS}
	After illustrating the physical meaning of sampling ZPE, we will now show that MMST dynamics can capture very accurate quantum dynamics for certain coupled electron-photonic systems. To begin with, we investigate the simplest scenario --- spontaneous emission for a single TLS in a cavity.
	
	For a TLS in a free vacuum ($L \rightarrow \infty$), the excited state population for the TLS decays exponentially with the Fermi's golden rule (FGR) rate:
	\begin{equation}
	\kFGR^{\text{1D}} = \frac{\omega_{0}\mu_{ge}^2}{\epsilon_0 \hbar c}
	\end{equation}
	In the cavity with finite $L$, the process is different. Before the EM field emitted by the TLS hits the cavity mirror, the TLS decays as in the free vacuum. However, due to the existence of the cavity mirror, the emitted EM field can reflect back and re-excite the TLS, known as a Poincar\'e recurrence.

	Fig. \ref{fig:SE_popu_MMST} plots the time evolution of the electronic populations for spontaneous emission of a single TLS in a 1D cavity with length $L = 2\pi$. The solid lines denote the excited state population and the dashed lines denote the ground state population. In general, MMST dynamics (red lines) predict almost the same population dynamics as QED (black lines): initially,
	 because the TLS locates at the center of the cavity, when $t < \pi$, the excited state population of the TLS exponentially decay with rate $\kFGR^{\text{1D}}$; later, at  $t = 2\pi$,  Poincar\'e recurrence occurs and the TLS is re-excited. 
	 \red{One interesting finding in Fig. \ref{fig:SE_popu_MMST} is that although MMST dynamics predict the initial populational decay relatively well, this approach considerably underestimates the height of the Poincar\'e recurrence. Such an underestimation must come from the independent-trajectory assumption in MMST dynamics.  MMST dynamics cannot capture all coherence effects and, for long times, such errors will necessarily be amplified. \blue{Throughout this paper, because we are interested in the dynamics for not very long times, MMST dynamics perform relatively well.}}

	Having  identified the physical meanings of  the ZPEs  in MMST dynamics, we can now use MMST dynamics to investigate the respective contribution of self-interaction (vacuum fluctuations) to both the initial decay and the Poincar\'e recurrence. As shown in Fig. \ref{fig:SE_popu_turnoff_samplings}, at the initial stages of spontaneous emission ($t \approx 0$), both self-interaction (blue \red{dashed}  line) and vacuum fluctuations (green \red{dash-dot}  line) contribute to spontaneous emission almost equally, \red{which agrees with DDC's calculation of spontaneous emission in free space\cite{Dalibard1982,Dalibard1984}}. After the initial exponential decay ($t = \pi$), just  as in Fig. \ref{fig:ground_state_popu}, consideration of only self-interaction leads to the breakdown of \red{the} electronic ground state, and consideration of only  vacuum fluctuations leads to spontaneous absorption.
	Finally and very interestingly, the MMST dynamics in Fig. \ref{fig:SE_popu_turnoff_samplings} inform us that a Poincar\'e recurrence ($t = 2\pi$) is caused mostly by self-interaction rather than vacuum fluctuations.

	Apart from electronic population dynamics, MMST dynamics can also capture accurately the evolution of the real-space distribution for the E-field intensity. In Fig. \ref{fig:SE_intensity_MMST}, we plot the  real-space distribution of the E-field intensity from time $t = 0$ to $t = 3\pi$ (from bottom to top) with time interval $0.3 \pi$. When $1.28\times 10^6$ trajectories are averaged, MMST dynamics (red \red{solid}  line)  agree with QED (black \red{dashed}  line) very well: Initially the  EM field is generated from the cavity center (where the TLS is located); at $t = \pi$, the emitted EM field reaches the cavity mirror; at $t = 2\pi$, the emitted EM field propagates back to the cavity center and re-excited the TLS; at later times, the excited TLS emits another small EM bump after the wavefront. The only difference between MMST dynamics and exact dynamics, however, is the non-vanishing intensity oscillations at the cavity center ranging from negative to positive values.
	\red{On the one hand, from our perspective, the \textit{negative} values of the MMST intensity oscillations in Fig. \ref{fig:SE_intensity_MMST} appear to be the result of numerical errors in the FDTD simulations. } 
	In practice, if we take a coarse-grained average (\red{cyan dash-dot} line) of the MMST results (red \red{solid}  line) by averaging over the neighboring 50 grids points (where we use 5001 grids points for the cavity), \red{the negative-value feature is}  completely smoothed away. \red{As shown in Appendix \ref{Appendix: middle_peak}, this negative feature also vanishes if the FDTD propagator is replaced by propagating $X_j$ and $P_j$ directly.}
	\red{On the other hand, by including the doubly excited state (while we use only the CIS approximation), quantum calculations actually predict a putatively similar non-vanishing middle peak\cite{Flick2017}, which would almost imply that the MMST results may contain more information than the CIS quantum results (which do not have a middle peak). And yet, because the size of these MMST  middle oscillations can depend sensitively on  \blue{the size of basis we use for propagating MMST dynamics}, our overall conclusion is that we should be very hesitant to attribute too much to these \textit{MMST} oscillations.} See  Appendix \ref{Appendix: middle_peak} for a detailed discussion for the "middle oscillations".

	Next, note that, for all of  results above, the TLS has been set at the center of the cavity ($r = L/2$). In QED, because the light-matter coupling term $g_j^{(\alpha)}$ is proportional to $\sin(k_j r_\alpha)$ (see Eq. \eqref{eq:gj}), the spontaneous emission rate will be strongly modified by changing the location of the TLS. For example, if the TLS is close to the cavity mirror with distance $r = \lambda/2$ (where $\lambda = 2\pi / \omega_{0}$ denotes the wavelength of the emitted EM field by the TLS), because the resonant cavity mode ($k_0 = \omega_{0}/c$) completely decouples to the TLS ($\sin(k_0 r) = 0$),  the spontaneous emission rate should be strongly inhibited. 
	Encouragingly, this scenario is correctly predicted by MMST dynamics.
	\red{On the left of Fig. \ref{fig:SE_near_mirror}, we plot the excited state population for a TLS when this TLS is initially set reasonably close to the left mirror with distance $\lambda/2$, where $\lambda=2\pi/\omega_{0}$ denotes the intrinsic wavelength for the TLS. When time $t<\lambda/2c$, the EM field emitted by the TLS has not yet  hit the left mirror, and so the TLS must evolve as in the free space (i.e., spontaneous decay with a Fermi's golden rule rate), which corresponds to the short quick decay in Fig. \ref{fig:SE_near_mirror}(left). However, when $t>\lambda/c$, because the reflected EM field returns to engage the TLS, the EM field felt by the TLS is actually an interference between the reflected EM field at earlier time ($t-\lambda/c$) and the emitted EM field at current time ($t$). Due to the distance to the mirror, such interference can be either constructive or destructive. For destructive interference, as shown in Fig.  \ref{fig:SE_near_mirror}(left), the excited state population decays much slower after $t=\lambda/c$. When the distance to the mirror is changed from $\lambda/2$ to $\lambda/4$, as shown in the Fig.  \ref{fig:SE_near_mirror}(right), the interference is constructive, and a faster-than-free-space decay can be observed.}

	\subsection{Modification of Spontaneous Emission for an Array of TLSs}
	After investigating spontaneous emission for a single TLS in the cavity, we now move to a more complicated case: an array of $N = 101$ equally spaced TLSs in the cavity, but we assume that only the middle TLS (which is located at the cavity center) is excited initially; this  scenario  can  be solved quantum-mechanically within the CIS ansatz. By changing the spacing $a$ between the TLSs, the population decay for the middle TLS can be significantly modified due to constructive or destructive interference between the emitted EM fields by TLSs. 
	
		\begin{figure*}
		\includegraphics[width=1.0\linewidth]{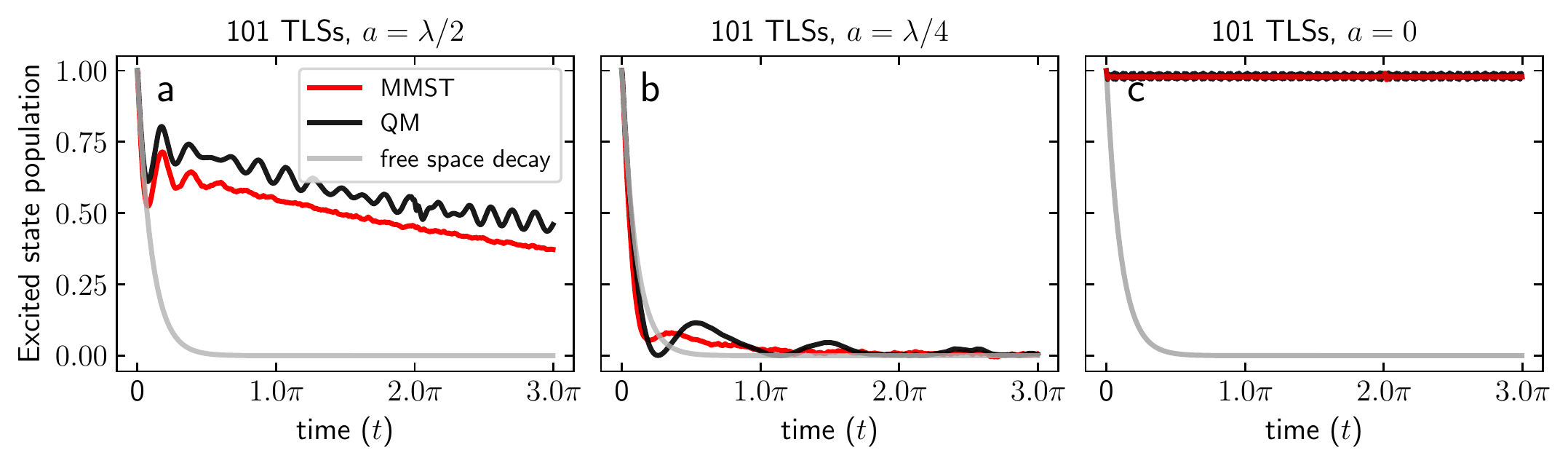}
		\caption{Time evolution of the excited state population for the excited TLS in the 1D cavity. For parameters, we set $N = 101$ identical TLSs equally spaced at the center part of the cavity and only the middle TLS is excited. From left to right, the spacing  between the neighboring TLSs is $a = \lambda/2$, $\lambda/4$, and 0, where $\lambda = 2\pi/\omega_{0}$. All other parameters are the same as Fig. \ref{fig:SE_popu_MMST}. Compared with the free space decay for a single TLS (gray line), MMST dynamics (red line) predict both enhancement and inhibition of the spontaneous emission rate when changing spacing $a$ in a similar way as the QED calculation (black line).}
		\label{fig:SE_chain}
	\end{figure*}
	
	Fig. \ref{fig:SE_chain} plots the population decay for the middle TLS with different spacing $a$. Overall,  MMST dynamics (red line) predict almost the same results as QED (black line):
	When $a = \lambda / 2$ (Fig. \ref{fig:SE_chain}a),  the spontaneous emission of the middle TLS is inhibited ---	the decay rate is smaller than $\kFGR^{\text{1D}}$ (gray line). By contrast, when  $a = \lambda / 4$ (Fig. \ref{fig:SE_chain}b), the spontaneous emission of the middle TLS is enhanced --- the decay rate is larger than $\kFGR^{\text{1D}}$. 
	Finally, when $a = 0$ (Fig. \ref{fig:SE_chain}c), the spontaneous emission of the middle TLS is strongly inhibited.
	Interestingly, the decay behavior in Figs. \ref{fig:SE_chain}a-b is very close to Fig. \ref{fig:SE_near_mirror}, suggesting that the effect of the surrounding TLSs is very similar to the effect of a cavity mirror\cite{Mirhosseini2019}.  \blue{However,  the intrinsic quantum mechanism for modifying the spontaneous emission rate in  Fig. \ref{fig:SE_chain} is different from Fig. \ref{fig:SE_near_mirror}: the rate modification in Fig. \ref{fig:SE_chain} comes from the different decay rates of the bright or the dark states formed by  $N$ TLSs, while the rate modification in Fig. \ref{fig:SE_near_mirror} comes from the interference with the reflected EM field.}

	\subsection{Dicke's Superradiance for a Collection of Excited TLSs}
	\red{In the previous two subsections,} only one TLS was excited initially, which can be easily propagated quantum-mechanically even when $N$ is large. By contrast, if we assume that all TLSs are excited initially, propagating the full (both electronic and photonic) quantum dynamics becomes  impossible for more than a few TLSs. One specific example of this scenario is Dicke's superradiance\cite{Dicke1954}, where all TLSs are located within one wavelength in the free vacuum. For Dicke's superradiance problem, due to the coherence between TLSs, the spontaneous emission rate for  a single TLS is proportional to the number of total TLSs $N$. 
	
		\begin{figure}
		\includegraphics[width=1.0\linewidth]{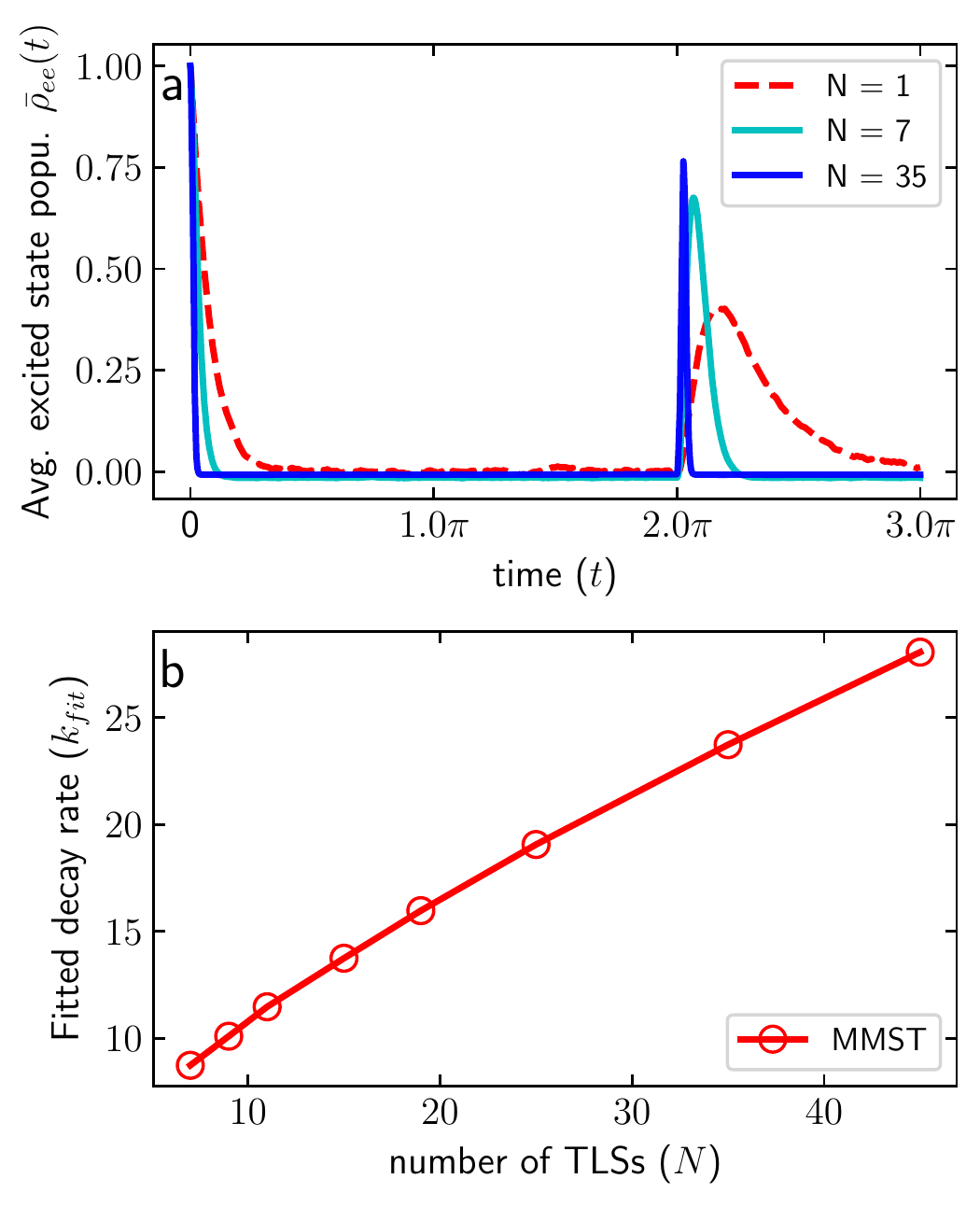}
		\caption{MMST dynamics for Dicke's superradiance: (a) time evolution of the averaged excited state population [$\bar{\rho}_{ee}(t)$] for the TLSs with $N = 1$ (red dashed), $7$ (\red{cyan} solid), and $35$ (blue solid) TLSs; (b) the fitted initial decay rate as a function of number of TLSs, where the red dots denote simulation data points. Here all $N$ TLSs are located at the center of the cavity ($r = L/2$) and start in the excited state. For each simulation,  the simulation time step is set to $\Delta t = \Delta x / 10c$, and all other parameters are the same as Fig. \ref{fig:SE_popu_MMST}. Note that the linear scaling of the fitted decay rate ($k_{\text{fit}}\propto N$) agrees with Dicke's prediction\cite{Dicke1954}. More interestingly, MMST dynamics also predict that when $N$ increases, the peak of the Poincar\'e recurrence is enhanced and the recurrence narrows in time.}
		\label{fig:SE_Dicke}
	\end{figure}
	
	Even though a full quantum simulation cannot be performed here, MMST dynamics can be easily propagated. To model Dicke's superradiance, here we assume that all TLSs are located at the center of the cavity and start in excited state. As shown in Fig. \ref{fig:SE_Dicke}a, with an increasing  number of TLSs, the initial exponential decay is accelerated. 
	Quantitatively speaking, by fitting the initial exponential decay to obtain an effective rate (i.e., supposing $\bar{\rho}_{ee}(t) = \exp(-k_{fit}t)$), we find that MMST dynamics can capture the linear scaling of the decay rate as a function of  $N$  correctly; see Fig. \ref{fig:SE_Dicke}b.
	More interestingly,  because this simulation is performed in cavities, MMST dynamics also predict a change in the Poincar\'e recurrence as a function of the number of TLSs $N$, i.e., when $N$ increases, the peak of the Poincar\'e recurrence is enhanced and the recurrence narrows.

	\begin{figure}
		\includegraphics[width=1.0\linewidth]{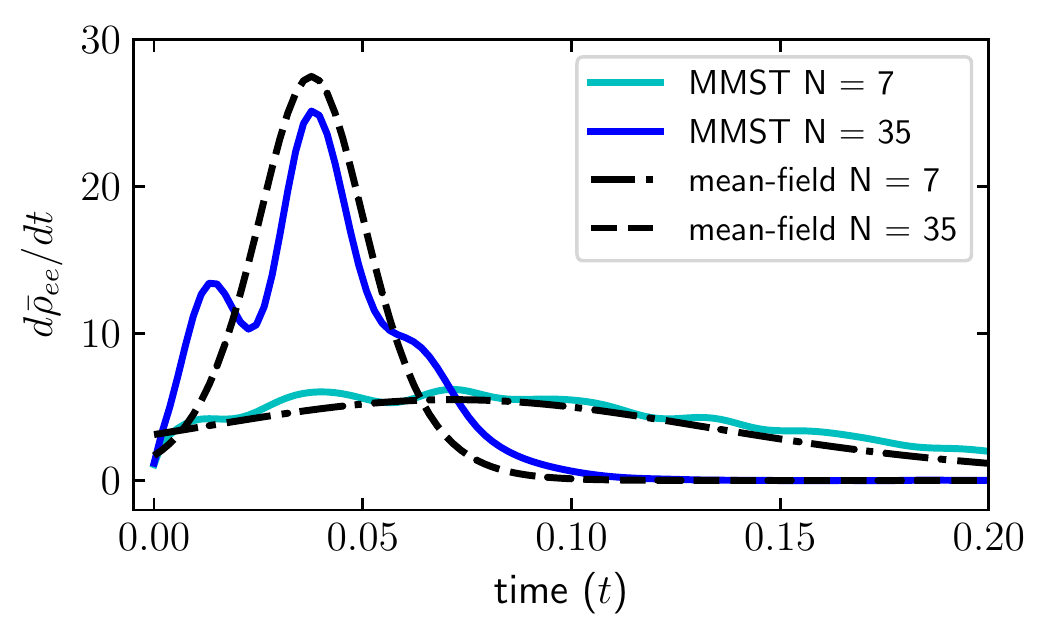}
		\caption{Time derivative of the averaged excited state population ($d\bar{\rho}_{ee}(t)/dt$) associated with Fig. \ref{fig:SE_Dicke} (left) during the initial population decay. Note that MMST dynamics (with classical EM fields) predict values for  $d\bar{\rho}_{ee}(t)/dt$ which are consistent with the mean-field  solution in Eq. \eqref{eq:drhodt_Dicke_sc} (black lines).}
		\label{fig:SE_Dicke_delay_time}
	\end{figure}

	Going beyond a simple rate expression, one might ask: do MMST dynamics predict the \red{correct} population dynamics during the  initial population decay?
	For example, according to a mean-field treatment of Dicke's superradiance,  the time derivative of the averaged excited state population $d\bar{\rho}_{ee}(t)/dt$ for $N$ TLSs obeys\cite{Gross1982}
	\begin{equation}\label{eq:drhodt_Dicke_sc}
	\frac{d\bar{\rho}_{ee}(t)}{dt}= \frac{\kFGR^{\text{1D}} N}{4}\left[\cosh\left(\frac{\kFGR^{\text{1D}} N}{2}\left(t - t_D\right)\right)\right]^{-2}
	\end{equation}
	where $t_D$
	denotes the delay time, at which $d\bar{\rho}_{ee}(t)/dt$ takes the maximum value. Eq. \eqref{eq:drhodt_Dicke_sc} not only shows that the spontaneous emission rate is proportional to $N$, but also implies that $\frac{d\bar{\rho}_{ee}(t)}{dt}$ has a burst at the delayed time $t_D$ (instead of at time zero). \red{Can this feature be captured by MMST dynamics?}
	
	We answer this question in Fig. \ref{fig:SE_Dicke_delay_time}. By plotting the time derivative of the averaged population dynamics ($d\bar{\rho}_{ee}(t)/dt$), we show that MMST dynamics \red{do} capture almost the same $d\bar{\rho}_{ee}(t)/dt$ as the mean-field expression in Eq. \eqref{eq:drhodt_Dicke_sc}   during the initial population decay, where the delay time ($t_D$) in Eq. \eqref{eq:drhodt_Dicke_sc} is taken from the MMST results.

	We note that Fig. \ref{fig:SE_Dicke_delay_time} already characterizes the basic features of superradiance, showing the usefulness of MMST dynamics. However, due to the quantum fluctuations of the initial phase of the superradiance process, the observed delay time exhibits some uncertainties from one experimental realization to another, 	 which can  not be predicted by the mean-field expression in Eq. \eqref{eq:drhodt_Dicke_sc}.  A quantum estimation suggests that the statistics of the delay time should obey\cite{Gross1982}:
		\begin{subequations}\label{eq:delay_time_statics}
		\begin{align}
		\avg{t_D} &\sim \frac{1}{N\kFGR^{\text{1D}}}\sum_{s=1}^{N} \frac{1}{s} \\
		\Delta t_D &= \sqrt{\avg{t_D^2} - \avg{t_D}^2} \sim \frac{1}{N\kFGR^{\text{1D}}}\sqrt{\sum_{s=1}^{N} \frac{1}{s^2}}
		\end{align}
	\end{subequations}

	 	\begin{figure}
	 	\includegraphics[width=1.0\linewidth]{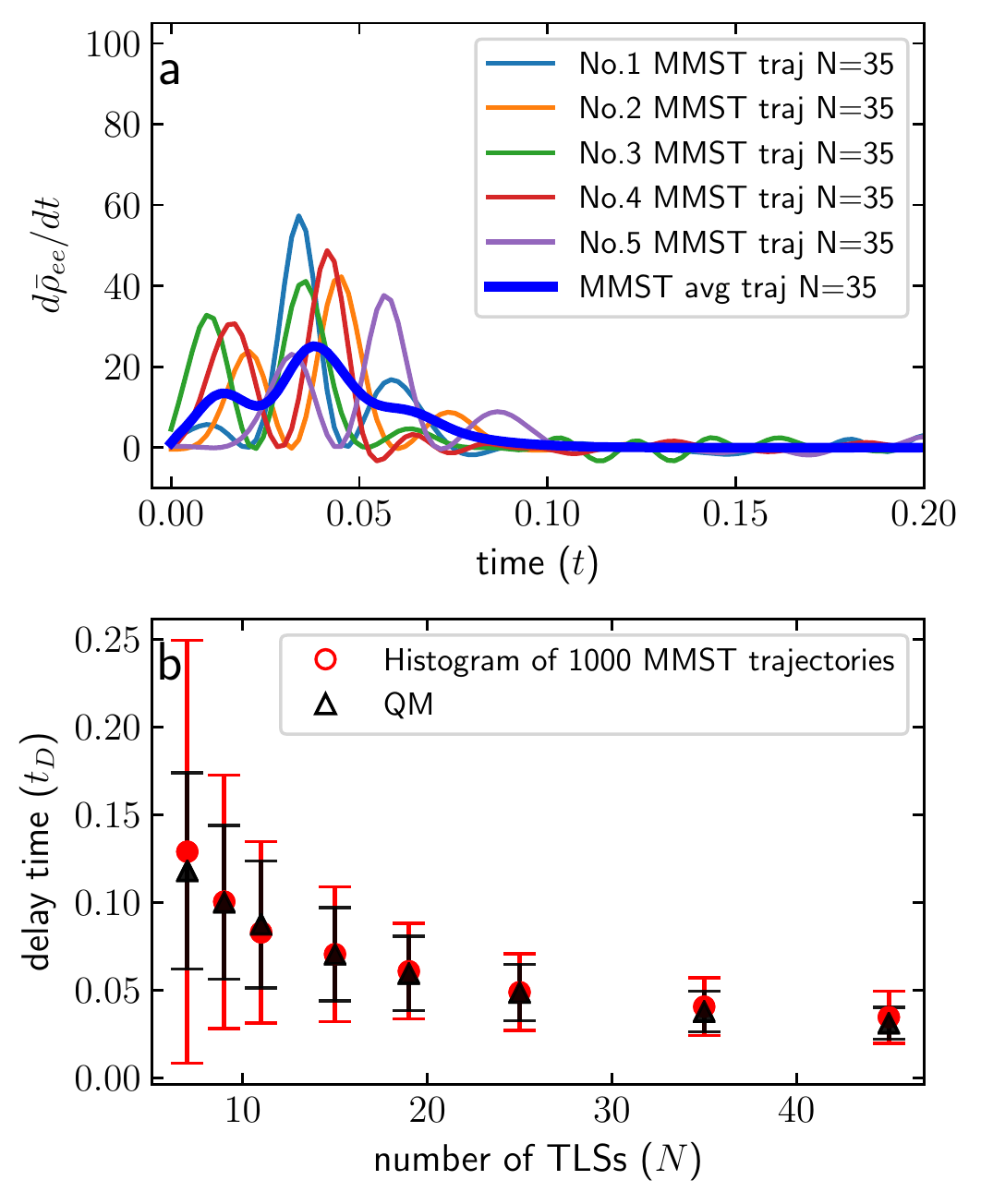}
	 	\caption{ Delay time statistics for Dicke's superradiance. (a) 
	 		Time evolution of  $d\bar{\rho}_{ee}(t)/dt$  for different MMST trajectories when $N=35$. There are five thin colored lines, each represents one MMST trajectory and the bold blue line denotes the averaged MMST trajectory as shown in Fig. \ref{fig:SE_Dicke_delay_time}. Note that the delay times for MMST trajectories exhibits strong fluctuations. (b) Statistics of the delay time for different number of TLSs. For MMST dynamics, when taking the statistics over 1000 trajectories, the expectation value (red dots) and standard deviation (red error bars) of the delay time  \red{agree surprisingly well with the quantum estimate} (black) as shown in Eq. \eqref{eq:delay_time_statics}. }
	 	\label{fig:SE_Dicke_trajs}
	 \end{figure}
 
 	 \begin{figure}
 	\includegraphics[width=1.0\linewidth]{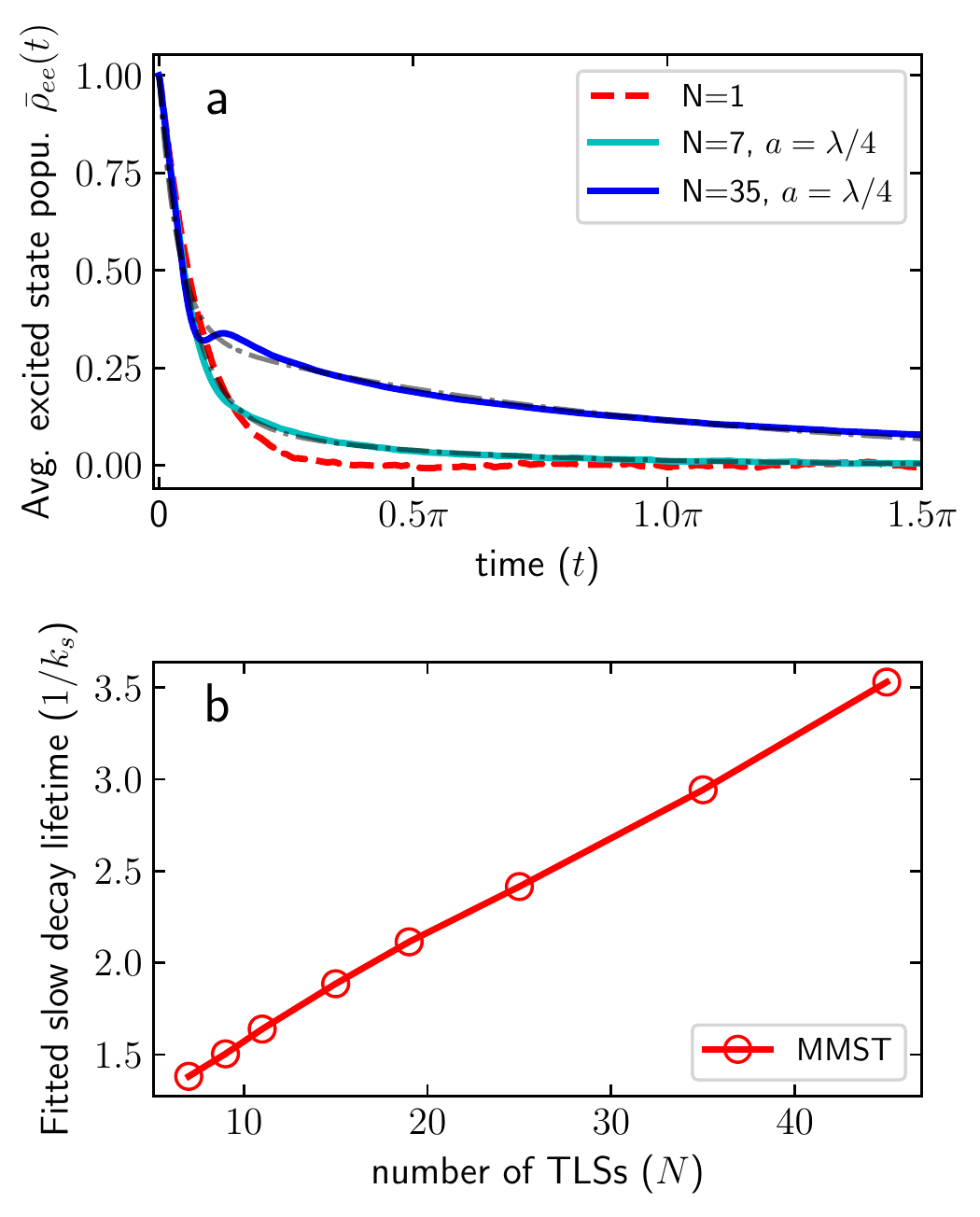}
 	\caption{Dicke's subradiance for an array of equally spaced TLSs with separation $a = \lambda/4$; all other parameters are same as in Fig. \ref{fig:SE_Dicke}. The system starts with all $N$ TLSs excited. (a) Time evolution of the averaged excited state population ($\bar{\rho}_{ee}(t)$) for the TLSs with $N = 1$ (red dashed), $7$ (\red{cyan} solid), and $35$ (blue solid) TLSs. The dash-dot lines denote the biexponential fit (i.e., $\bar{\rho}_{ee}(t) = A\exp(-k_s t) + (1-A)\exp(-k_f t)$, where $k_s$ and $k_f$ denote the slow and fast decay rates). (b) The fitted slow decay lifetime ($1 / k_s$) as a function of $N$.  Note that the subradiant lifetime displays a linear scaling  as a function of $N$ ($1/k_s\propto N$).}
 	\label{fig:SE_Dicke_subradiance}
 \end{figure}
	 
	 At this point, keen readers might ask: can MMST dynamics also predict \red{such} statistics of the delay time for superradiance?  Indeed, by further investigating the dynamics of $d\bar{\rho}_{ee}(t)/dt$ for different MMST trajectories, in Fig. \ref{fig:SE_Dicke_trajs}a we show that the delay times for different MMST trajectories (thin colored lines) exhibit strong fluctuations. The fluctuations of the delay time of MMST trajectories reflect dynamics beyond the mean-field description in Eq. \eqref{eq:drhodt_Dicke_sc}. 
	 Most interestingly, as shown in Fig. \ref{fig:SE_Dicke_trajs}b, by plotting the statistics of the delay times for 1000 MMST trajectories\red{,   we find that} the expectation value (red dots) and the standard deviation (red error bars) of the delay time agrees surprisingly well with the quantum estimation (black, see Eq. \eqref{eq:delay_time_statics}).

	 After demonstrating  Dicke's superradiance for \red{a  sample of TLSs at a single point in space}, we now consider a different limit: when the separation ($a$) between TLSs is comparable with the emitted wavelength ($\lambda$). In this limit, because the  dark states can also couple to the environment,  the collective emission behavior is no longer superradiant. \red{For} an array of $N$ equally spaced TLSs with separation $a = \lambda/4$ for example, Fig. \ref{fig:SE_Dicke_subradiance}a shows the MMST results for the averaged excited state population ($\bar{\rho}_{ee}(t)$) as a function of time. For $N>1$, the decay behavior is indeed no longer exponential: after a fast initial decay, $\bar{\rho}_{ee}(t)$ demonstrates a slow decay in the long times. By further fitting $\bar{\rho}_{ee}(t)$ with a biexponential function (\red{i.e.,} $A\exp(-k_s t) + (1-A)\exp(-k_f t)$, where $k_s$ and $k_f$ denote the slow and fast decay rates), we extract the slow decay rate $k_s$ for different $N$. As shown in Fig. \ref{fig:SE_Dicke_subradiance}b, the subradiant decay lifetime ($1/k_s$) exhibits near linear scaling as a function of $N$, which agrees with recent cold-atom experiments\cite{Guerin2016,Weiss2019}.

	\section{Discussion and Conclusion}\label{sec:disscussion_conclusion}
	In summary, we have used MMST dynamics to solve a specific coupled electron-photonic system in cQED --- a collection of $N$ TLSs coupled to a multimode cavity. 
	The spirit of MMST dynamics is to approximate  quantum dynamics by sampling independent quasiclassical trajectories in  Wigner phase space,  the initial conditions of which are determined by sampling both the electronic and photonic ZPEs.
	\red{MMST dynamics are ideal for electron-photonic systems, because photons are intrinsically harmonic. Because
	 propagating \textit{Cartesian} coordinates  is unnecessarily cumbersome, we have chosen to propagate the coupled Maxwell-Schr\"odinger equations via a split-operator-finite-difference-time-domain (SO-FDTD) propagator in order to reduce the computational cost for each trajectory.}
	
	Armed with the appropriate subroutine for MMST dynamics,
	we find that this algorithm can provide an intuitive and practical way to identify the respective contributions of self-interaction and vacuum fluctuations: sampling electronic ZPE reflects radiative self-interaction and sampling photonic ZPE reflects vacuum fluctuations. 
	\red{Near the ground state, these two effects balance each other, and so it is perhaps not surprising that traditional mean-field (Ehrenfest)  dynamics (without any ZPE) become accurate in this limit\cite{Crisp1969,Li2018Tradeoff,Li2019Hamiltonians,Chen2019Mollow}.  By contrast, at high saturation limit, while traditional mean-field (Ehrenfest) dynamics usually fail, MMST dynamics can still recover accurate quantum dynamics, at least for the test cases studied here:}
	(i) MMST dynamics accurately capture the initial exponential decay, the Poincar\'e recurrence, and the position dependence of the  spontaneous emission rate for a TLS in a cavity.
	(ii) For an array of $N  =101$ equally spaced TLSs in a 1D cavity and with only the middle TLS excited initially, MMST dynamics predict the modification of exponential decay (i.e., enhancement  and inhibition) accurately as a function of the spacing between TLSs. 
	(iii) MMST dynamics can model Dicke's superradiance and subradiance (i.e., when all TLSs are excited and located within one wavelength) and  correctly predict  the quantum statistics of the delay time.
	
	With these exciting findings of MMST dynamics in mind, we end this paper with several questions, which need to be answered for future improvement:
	
	(i) First, note that the computational cost of MMST dynamics is relative expensive: it usually requires $10^4-10^6$ trajectories for convergent results, thus preventing the application of MMST dynamics to more realistic systems (e.g., a three-dimensional cavity). Therefore, our first question is: are there enhanced sampling techniques for recovering the same-level MMST results with  fewer trajectories?
	
	(ii) Second, currently MMST dynamics require sampling a lot of cavity photon modes. When considering a free space problem, however, in principle we need to sample infinitely many modes, i.e. a continuous number of  photon modes to avoid any potential Poincar\'e recurrences, and sampling so many modes will be difficult in practice. Is there a better way to generalize MMST dynamics to  free space?
	
	(iii) Third, in general, quasiclassical MMST (or LSC-IVR) dynamics are known not to recover the correct equilibrium distribution, or even obtain detailed balance for a given quantum subsystem\cite{Habershon2009}.  In general, ZPE leakage is a  big problem\cite{Muller1999,Habershon2009}.  However, the electron-photonic case is potentially special because the photon bath is exclusively harmonic, whereas ZPE leakage is normally associated with anharmonic modes. Moreover, with harmonic surfaces, one never encounters negative forces\cite{Bellonzi2016}  that can also lead to unstable MMST trajectories.  Nevertheless, Hoffmann \textit{et al} have recently shown that ZPE leakage does become a problem with three-state electronic models (rather than two-state models)\cite{Hoffmann2019Benchmark}.  Thus, one must inquire: might SQC dynamics\cite{Cotton2013,Cotton2016,Cotton2019} and similar approaches provide better results in this limit? 
	
	(v) Fourth, MMST dynamics have (to date) usually been applied to  coupled electron-nuclear systems; can we directly apply MMST dynamics to  coupled electron-nuclear-photonic systems for simulating realistic systems\cite{Orel1979}?
	
	There is a great deal of work to be done at the intersection of semiclassical dynamics and quantum electrodynamics.

		\section{Acknowledgement}\label{sec:acknowledgement}
	This material is based upon work supported by the U.S. Department of Energy, Office of Science, Office of Basic Energy Sciences under Award Number DE-SC0019397.
	\red{T.E.L. also acknowledges the Vagelos Institute for Energy Science and Technology at the University of Pennsylvania for a graduate fellowship.}
	The research of A.N. is supported by the Israel-U.S. Binational
	Science Foundation.
	This research also used resources of the National Energy Research Scientific Computing Center (NERSC), a U.S. Department of Energy Office of Science User Facility operated under Contract No. DE-AC02-05CH11231.

		\begin{appendices}

		\section{The Middle Oscillations in Fig. \ref{fig:SE_intensity_MMST}: Numerical Error or Polariton?}
		\label{Appendix: middle_peak}
		
		\setcounter{equation}{0}
		\setcounter{figure}{0}
		\setcounter{table}{0}
		\renewcommand{\theequation}{A\arabic{equation}}
		\renewcommand{\thefigure}{A\arabic{figure}}
		
		\begin{figure*}
			\includegraphics[width=0.9\linewidth]{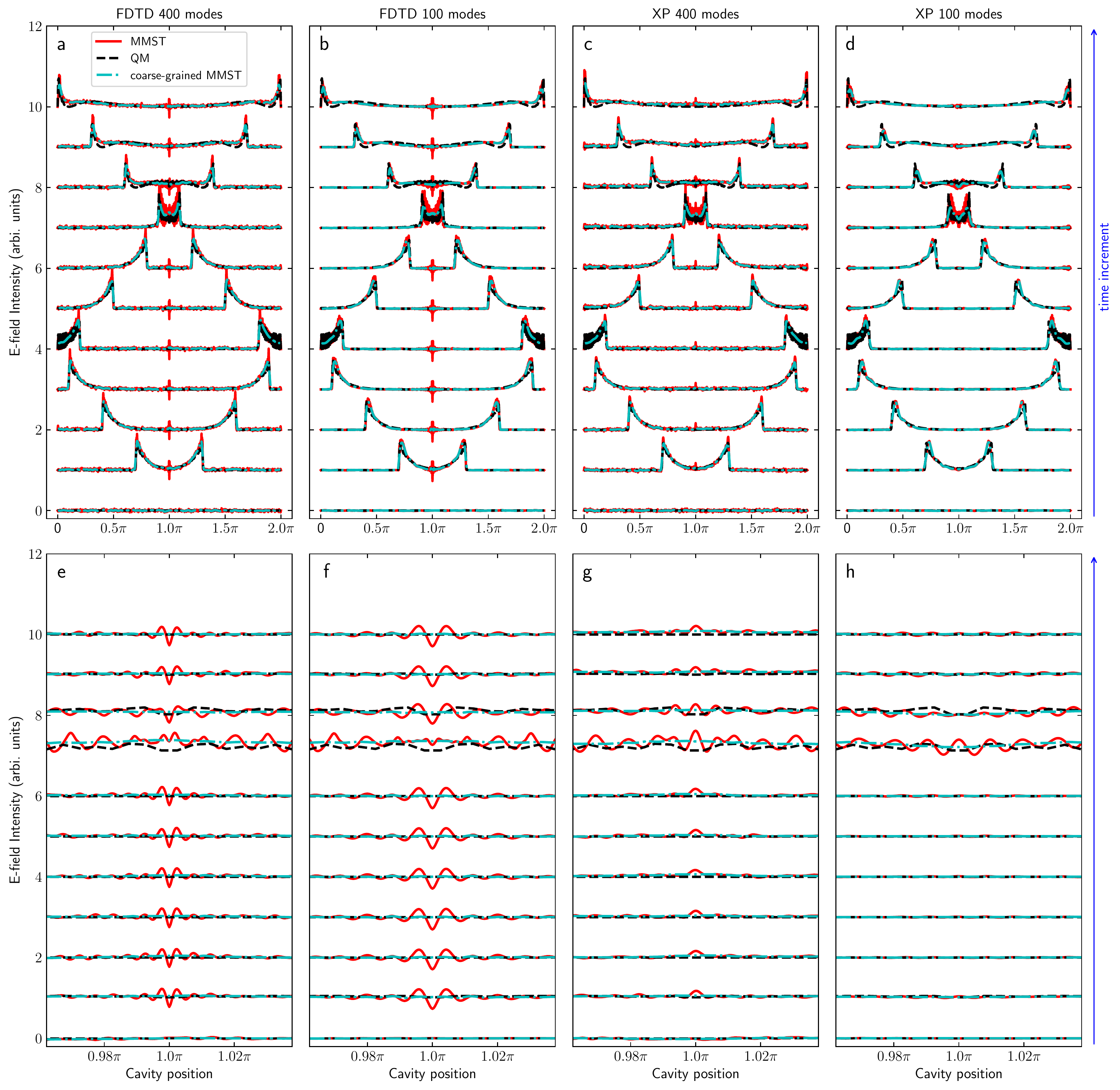}
			\caption{Real-space distribution of the E-field intensity as predicted by four different numerical treatments: (a) a replot of Fig. \ref{fig:SE_intensity_MMST}, i.e., initializing the EM field with 400 photon modes and propagating the EM field with FDTD; (b) the same as Fig. \ref{fig:EM_compare}-a \red{but the initialization is with} only 100 photon modes (centered at $\omega_{0}$); (c) initializing  400 photon modes and propagating $\{X_j, P_j\}$ directly; (d) the same as Fig. \ref{fig:EM_compare}-c but initializing only 100 photon modes. The bottom subplots are the corresponding zoom-in near the cavity center corresponding to the top zoom-out subplots. Note that the \red{negative-value feature of the} "middle oscillations" in Figs. \ref{fig:EM_compare}-a,b appears to be the result of numerical error with FDTD, and the middle MMST peak in Fig. \ref{fig:EM_compare}-c disappears with fewer photon modes.} 
			\label{fig:EM_compare}
		\end{figure*}
	
		When considering Fig. \ref{fig:SE_intensity_MMST}, note that a few references have reported a similarly non-vanishing peak in the \textit{E} field located at the middle of a 1D cavity (rather than \red{the oscillations in Fig. \ref{fig:SE_intensity_MMST}}) when modeling spontaneous emission and the claim has been made that this peak represents a polariton\cite{Flick2017,Hoffmann2019}.
		\red{Now}, guided by causality, one would presume that before the emitted EM field hits the cavity mirrors, the TLS should behave in a cavity the same as in free space\cite{Meystre2007}. 
		\red{Therefore, the existence of a middle peak in Refs. \cite{Flick2017,Hoffmann2019} (or the middle oscillations in Fig. \ref{fig:SE_intensity_MMST}) arise(s) two obvious questions: (i) Is the middle peak in Refs. \cite{Flick2017,Hoffmann2019} real  or the result of numerical errors in a quantum calculation? (ii) Are the middle oscillations in Fig. \ref{fig:SE_intensity_MMST} real  or numerical errors in MMST dynamics? Our current belief is that these two questions need not be linked.
		
		For the first question, our quantum simulation cannot (unfortunately) provide a  definite answer. 	 In our simulations, the use of the CIS basis  implies that all counter rotating-wave terms for the light-matter coupling have been ignored. In order to investigate whether the middle peak is real or not, however, one would need to consider the effect of such counter rotating-wave terms and use (at least) the two-photon states (i.e., the CISD approximation).
		For the results presented in Refs. \cite{Flick2017,Hoffmann2019}, the CISD approximation is used, and the middle peak is observed. Hence, one must presume  that the middle peak is real and comes from the effect of counter rotating-wave terms.

		For the second question, because MMST dynamics  do not invoke the rotating-wave approximation, this approach has the potential to predict the middle peak correctly, and the putative similarity between the quantum middle peak and the MMST middle oscillations might even  indicate that MMST dynamics can also predict this interesting quantum feature.
		However, as shown in Fig. \ref{fig:SE_intensity_MMST}, the middle oscillations in MMST dynamics contain negative values, which is not encouraging. To that end, let us now investigate the origin of this negative-value feature in more detail.

		 For all results presented above, we initialized the EM field with $M=400$ photon modes and propagated the EM field by the FDTD algorithm; see Sec. \ref{sec:simulation_details} for details. This numerical treatment leads to the E-field distribution shown in Fig. \ref{fig:SE_intensity_MMST}. To facilitate our discussion, here we will replot Fig. \ref{fig:SE_intensity_MMST}  in Fig. \ref{fig:EM_compare}-a; in Fig. \ref{fig:EM_compare}-e we zoom in near the cavity center.  
		
		In order to understand the origin of the middle oscillations,
		we have now run} the following additional simulations: (i) We initialize the EM field with fewer photon modes (say, $M=100$ modes centered at $\omega_{0}$ with frequency spacing $\Delta \omega = 0.5$) but still propagate the EM field with FDTD; in this case, the middle oscillations do not disappear. See Figs. \ref{fig:EM_compare}-b,f. (ii) As reported in Refs. \cite{Hoffmann2019,Hoffmann2019Benchmark}, if we directly propagate $M = 400$ photon modes  (i.e., propagating $\{X_j, P_j\}$) --- rather than running FDTD --- the middle oscillations are replaced by a non-vanishing middle peak, \red{which is close to the quantum result with a  CISD basis.} See		 \ref{fig:EM_compare}-c,g. (iii) Finally, if only  $M=100$ photon modes \red{(centered at $\omega_{0}$)} are propagated directly (i.e., not with FDTD), the middle peak \red{almost} completely disappears; see Figs. \ref{fig:EM_compare}-d,h.
		
		From  the simulations above, we must conclude that the the 
		\red{negative-value feature of the middle oscillations in Fig. \ref{fig:SE_intensity_MMST} arises}
		from the numerical errors with FDTD.  \red{Because reducing the number of the off-resonant photon modes leads to the disappearance of the middle peak when propagating $X_j$ and $P_j$, it appears  that  MMST dynamics cannot predict the middle peak in a reliable and consistent fashion.}
		As an alternative, the peak may arise from the ZPE leakage  from MMST dynamics\cite{Muller1999,Habershon2009}. \red{More analysis of this feature will be needed in the future.}

		\end{appendices}

		
		%

\end{document}